\documentclass[12pt,twoside,a4paper]{article}
\usepackage[dvips]{epsfig}
\newcommand{\simgt}{\,\rlap{\lower 3.5 pt \hbox{$\mathchar \sim$}}
  \raise 1pt \hbox {$>$}\,}
\newcommand{\simlt}{\,\rlap{\lower 3.5 pt \hbox{$\mathchar \sim$}}
  \raise 1pt \hbox {$<$}\,}
\begin{document}
\thispagestyle{empty} 
\title{
\vskip-3cm
{\baselineskip14pt
\centerline{\normalsize DESY 00--039 \hfill ISSN 0418--9833}
\centerline{\normalsize MZ-TH/00--06 \hfill} 
\centerline{\normalsize TTP00--02 \hfill}
\centerline{\normalsize hep--ph/0003082 \hfill} 
\centerline{\normalsize March 2000 \hfill}} 
\vskip1.5cm
Photon Plus Jet Cross Sections in Deep Inelastic \\ 
$ep$ Collisions at Order $O(\alpha^2 \alpha_s)$
\author{A.~Gehrmann-De Ridder$^1$, G.~Kramer$^2$ and H.~Spiesberger$^3$
\vspace{2mm} \\
{\normalsize $^1$ Institut f\"ur Theoretische Teilchenphysik,
  Universit\"at Karlsruhe,}\\ 
{\normalsize D-76128 Karlsruhe, Germany} \vspace{2mm}\\
{\normalsize $^2$ II. Institut f\"ur Theoretische
  Physik\thanks{Supported by Bundesministerium f\"ur Bildung und
    Forschung, Bonn, Germany, under Contract 05~HT9GUA~3, and
    by EU Fourth Framework Program {\it Training and Mobility of
    Researchers} through Network {\it Quantum Chromodynamics and
    Deep Structure of Elementary Particles}
    under Contract ERB FMRX--CT98--0194.}, Universit\"at
  Hamburg,}\\ 
\normalsize{Luruper Chaussee 149, D-22761 Hamburg, Germany} \vspace{2mm}
\\ 
\normalsize{$^3$ Institut f\"ur Physik,
  Johannes-Gutenberg-Universit\"at,}\\ 
\normalsize{Staudinger Weg 7, D-55099 Mainz, Germany} \vspace{2mm} \\
\normalsize{e-mail: gehra@particle.physik.uni-karlsruhe.de,} \\
\normalsize{kramer@mail.desy.de, hspiesb@thep.physik.uni-mainz.de}
} }
\date{}
\maketitle
\begin{abstract}
\medskip
\noindent
The production of a hard and isolated photon accompanied by one or two
jets in large-$Q^2$ deep inelastic $ep$ scattering is calculated at 
next-to-leading order. We include consistently contributions from
quark-to-photon fragmentation and study various differential cross
sections and their dependence on isolation cut parameters. Numerical
results relevant for HERA experiments are presented.
\end{abstract}

\clearpage
\section{Introduction}
In the past, measurements of prompt photon production at both
fixed-target facilities and hadron-hadron colliders, have extensively
been used to constrain the gluon distribution of the proton
\cite{apana}. Only recently the first data on prompt photon production
in high-energy $ep$ collisions have been reported \cite{HERA-data}. Due
to the presently still limited statistics the measurements are confined
to prompt photons produced in photoproduction reactions, i.e.\ to $ep$
collisions with almost real exchanged photons ($Q^2 \simeq 0$). As is
well-known, photoproduction processes at high energies proceed by two
distinct mechanisms. The incoming photon can couple either in a
point-like manner to the incoming quark or antiquark (direct process) or
hadronically as a source of quarks and gluons which in turn take part in
the subsequent hard scattering process (resolved process). Therefore an
important advantage of photoproduction measurements is to provide
additional constraints on the quark and gluon content of the photon as
suggested many years ago by Aurenche et al.\ \cite{aurenche}.

By contrast, prompt photon production in deep inelastic scattering with
large $Q^2$, $Q^2 \simgt 10$ GeV$^2$, is fully determined by the direct
process and does not need any non-perturbative input for the parton
content of the photon\footnote{At intermediate $Q^2 < 10$ GeV$^2$ there
  might be a significant resolved contribution which depends on the
  quark and gluon distribution of the {\em virtual} photon.} as in
photoproduction. Therefore this process is sensitive only to the parton
distribution functions (PDF's) of the proton. The possible information
on the proton PDF's would be complementary to the $F_2$ measurement from
inclusive deep inelastic scattering, since {\em up}- and {\em down}-type
quarks contribute with different weights. Of course, the cross sections
for $ep \rightarrow e\gamma X$ at large $Q^2$ are smaller than the
corresponding cross sections for almost real photons; but with the
larger luminosities planned at HERA rather accurate measurements of
various differential cross sections for $Q^2 > 10$ GeV$^2$ seem
feasible.

In addition to the perturbative direct production, photons are also
produced through the ``fragmentation'' of a hadronic jet into a single
photon carrying a large fraction of the jet energy \cite{koller}. This
long-distance process is described in terms of the quark-to-photon and
gluon-to-photon fragmentation functions (FF's).

In order to unambiguously identify the prompt photon signal from the
hadronic background it is necessary to apply isolation cuts in the
experiment. This has the effect of reducing the cross section due to a
suppression of the photon fragmentation contributions. On the other
hand, it has the advantage of eliminating to a large extent the
dependence on the photon fragmentation function, which is a
non-perturbative input and must come from other experiments designed to
measure them.

To obtain reliable predictions it is necessary to calculate the $ep
\rightarrow e\gamma X$ cross section in next-to-leading order (NLO) of
the strong coupling constant $\alpha_s$ as has been done for $p\bar{p}$
collisions \cite{ppbar}, $e^+e^-$ annihilation \cite{e+e-,GGdR1,GGdR2},
as well as photoproduction \cite{phopro}. The corresponding NLO
calculation of cross sections for prompt photon production in $ep$
scattering with large $Q^2$ has not yet been done previously, neither
for the technically simpler case of inclusive cross sections, i.e.\ 
without any photon isolation cut, nor for the case with isolation cuts.
Applicability of perturbative QCD requires that the scattering process
is characterized by a large transverse momentum, provided either by the
momentum transfer $Q^2$ or a large transverse momentum of the hadronic
final state. We consider only the case where both $Q^2$ is large and the
hadronic final state is characterized by a large $p_T$.  One specific
possibility is to consider the case where in addition to the photon also
one or more jets are observed in the final state. The detection of an
additional jet may also help to identify the prompt photon events in the
actual experiment.  In leading order (LO) the photon is produced by the
Compton process $\gamma^{\ast} + q \rightarrow \gamma + q$, where
$\gamma^{\ast}$ is a photon of high virtuality emitted by the incoming
electron. This partonic photon production process contributes to the
$\gamma +(1+1)$-jet final state in $ep$ scattering (the proton remnant
jet being counted as ``$+1$'' jet as usual). In NLO the final states are
$\gamma + (1+1)$- and $\gamma + (2+1)$-jets.

The first NLO calculations for this case were done by two of us and D.\ 
Michelsen \cite{KMS}. This calculation was restricted to the case of not
too large $Q^2$ where it is possible to neglect the exchange of a $Z$
boson. Moreover, in this previous work the fragmentation contribution
was not taken into account. Therefore, photon-quark collinear
singularities could not be absorbed into the fragmentation functions.
Instead, these singularities had been removed by explicit parton-level
cutoffs. As a consequence, the result depended strongly on these
parton-photon cutoffs, in particular for subprocesses with an incoming
gluon. These cutoffs are difficult to control experimentally, where
hadrons combined into jets are observed and not the partons needed to
define the cutoffs. In subsequent work \cite{GKS} we included the
fragmentation contribution thereby avoiding the need to use parton-level
cutoffs. The isolation criteria, which limit the hadronic energy in the
jet containing the photon, are thus physical, i.e.\ correspond to
selection criteria in the experimental analysis. In a later paper
\cite{GKS2} we studied the sum of the $\gamma + (1+1)$-jet and $\gamma +
(2+1)$-jet cross sections as a function of the momentum fraction carried
by a photon inside a jet. We observed that this special cross section is
sensitive to the fragmentation contribution, in particular to the
quark-to-photon FF.

In \cite{GKS}, only a few observables have been calculated, as for
example distributions with respect to the transverse momentum, $p_T$,
and the rapidity, $\eta$, of the photon or the most energetic jet for
one particular choice of the photon isolation cut. In this paper we take
up the topic of this earlier work. Besides several other observables
which are of interest for analyzing upcoming experimental data from HERA
we shall present results for $p_T$ and $\eta$ distributions already
considered in \cite{GKS} for different isolation cuts. We also study
possible scale dependencies to estimate the reliability of our
predictions. As in \cite{GKS} we use the $\gamma^{\ast}p$ center-of-mass
system to define the kinematic variables. For most of the cross section
calculations a particular cone algorithm is applied to define the parton
jets and to isolate the photon signal. For a few cases we shall also
make use of the $k_T$ cluster algorithm.

The plan of the paper is as follows. In section 2 a brief outline of the
theoretical background to the cross section calculations as well as the
technique of the calculation are given. The results for the various
observables are presented and discussed in section 3. Section 4 contains
a summary and some concluding remarks.

\section{Subprocesses Through Next-to-Leading Order}

\subsection{Leading-Order Subprocesses}

\begin{figure}[bh] 
\unitlength 1mm
\begin{picture}(160,40)
\put(31,-1){\epsfig{file=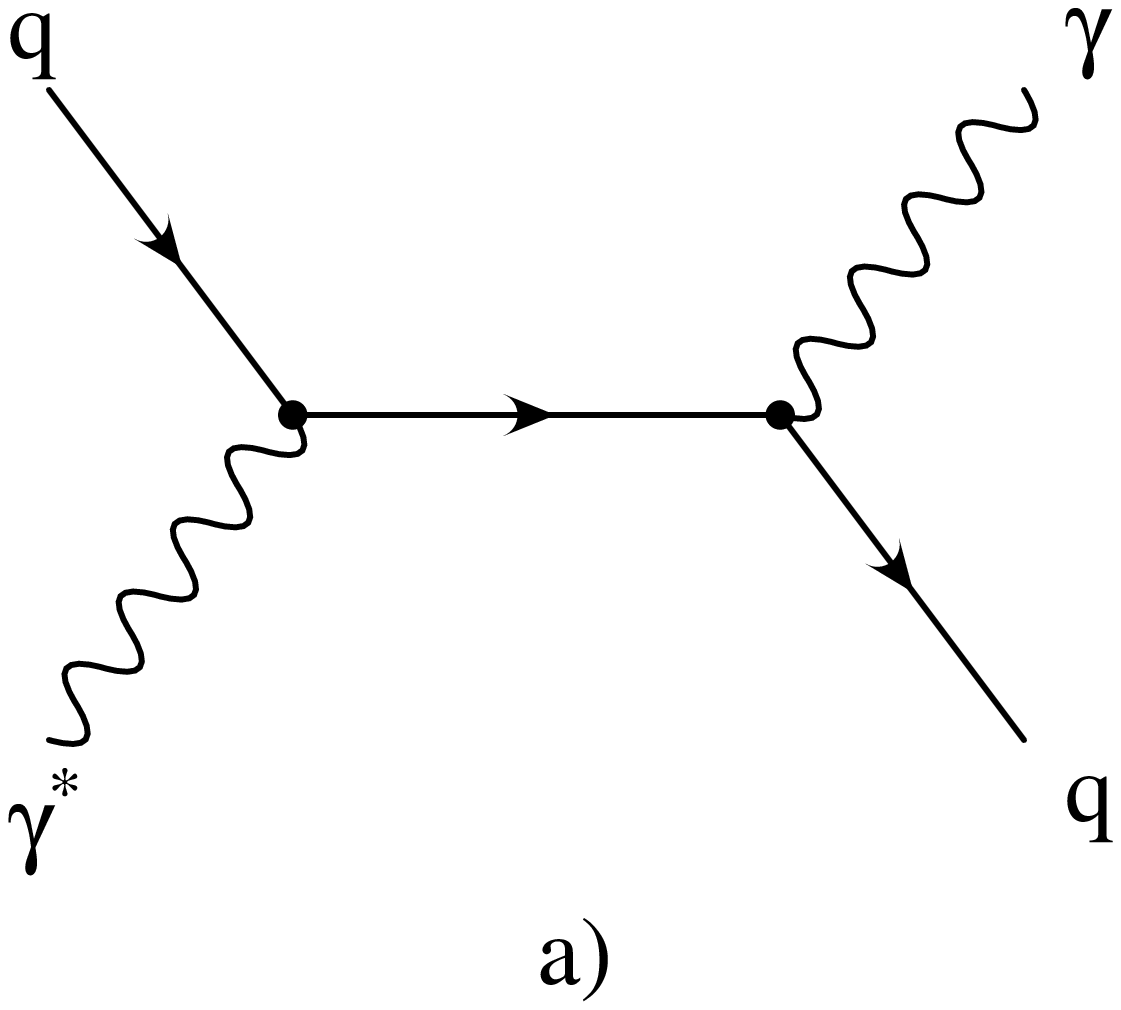,width=4cm}}
\put(91,-1){\epsfig{file=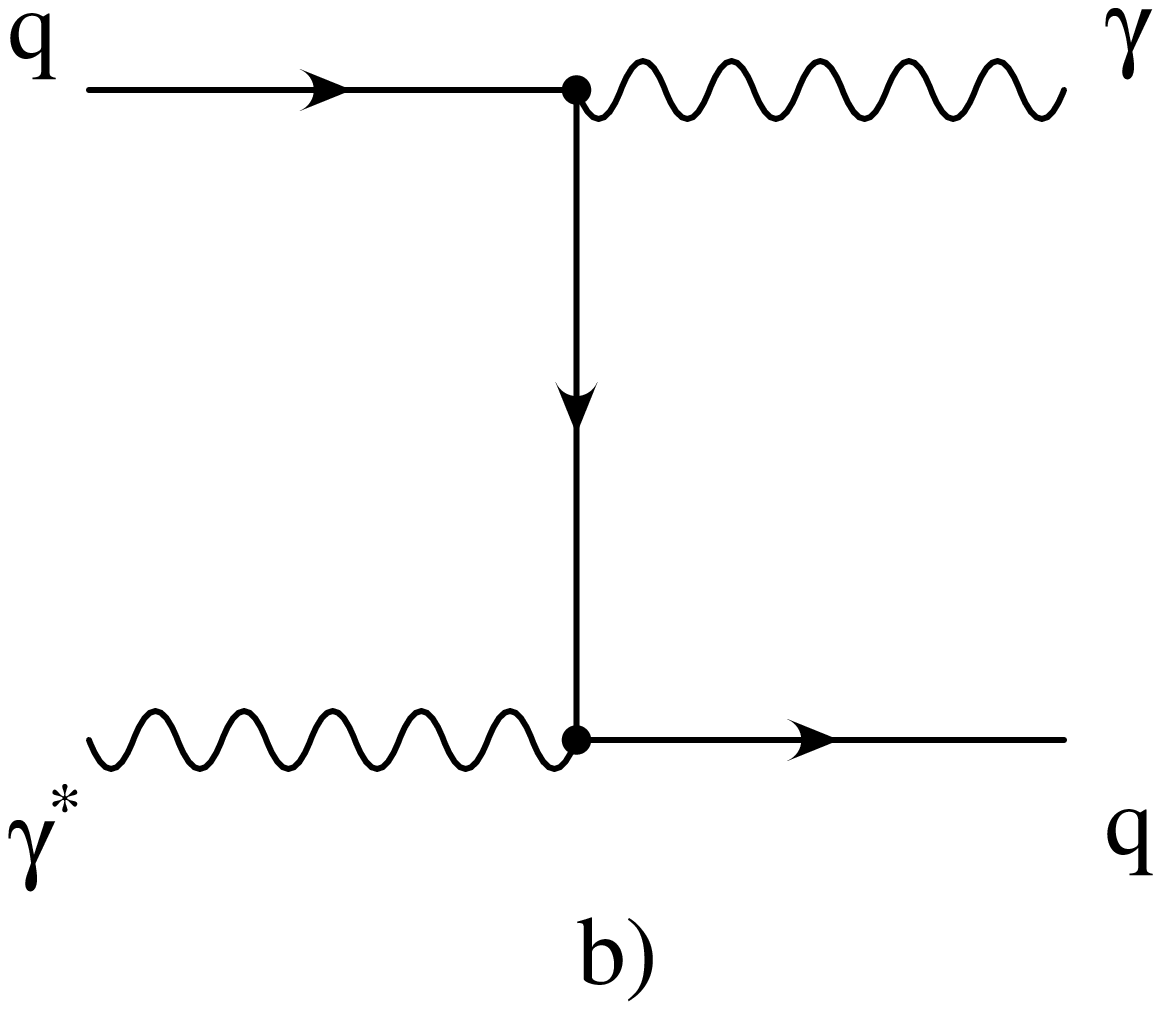,width=4cm}}
\end{picture}
\caption{Feynman diagrams for $\gamma^{\ast} + q \rightarrow q +
  \gamma$.} 
\label{fig1}
\end{figure}

In leading order, the production of photons in deep inelastic electron
scattering is described by the quark (antiquark) subprocess (see Fig.\ 
\ref{fig1})
\begin{equation}
e(p_1) + q(p_3) \rightarrow e(p_2) + q(p_4) + \gamma(p_5)
\label{loprocess}
\end{equation}
where the particle momenta are given in parentheses. The momentum of the
incoming quark is a fraction $\xi$ of the proton momentum $P$, $p_3 =
\xi P$. The proton remnant $r$ has the momentum $p_r = (1 - \xi) P$. It
hadronizes into the remnant jet so that the process (\ref{loprocess})
gives rise to $\gamma + (1+1)$-jet final states. In the virtual
photon-proton center-of-mass system the hard photon recoils against the
hard jet back-to-back. In a leading-logarithmic calculation, the effects
of higher-order processes show up only via the use of the
scale-violating parton distributions of the proton. The PDF's are
calculated using collinear kinematics so that the event structure is the
same as given by the lowest-order subprocesses, which are of order
$O(\alpha^2)$ {}\footnote{Here and in the following we do not count the
  extra factor $\alpha$ from the $ee \gamma^{\ast}$ vertex.}.

To remove photon production by incoming photons with small virtuality
and to restrict to the case where the scattered electron $e(p_2)$ is
observed, one applies cuts on the usual deep inelastic scattering
variables $x$, $y$, $Q^2$ as measured from the momentum of the scattered
lepton. In particular we restrict the calculation to values of $Q^2 \ge
10$ GeV$^2$; however, since very large $Q^2$ values are not relevant at
HERA we can neglect contributions from $Z$ boson exchange. In addition,
to have photons $\gamma(p_5)$ of sufficiently large energy we require an
explicit cut on their transverse momentum.  Finally, a cut on the
invariant mass of the hadronic final state is also applied.

Both leptons and quarks emit photons. The subset of diagrams where the
photon is emitted from the initial or final state lepton (leptonic
radiation) is explicitly gauge invariant and can be considered
separately. Similarly, the contribution from diagrams with a photon
emitted from quark lines is called quarkonic radiation. In addition,
there are also contributions from the interference of these two
mechanisms. The emission of photons from leptons is described by pure
QED and can be predicted with high reliability. Therefore, the
contributions from leptonic radiation will be suppressed by cuts on the
photon emission angle \cite{KMS}.  In our numerical evaluation we
include the remaining background from leptonic radiation as well as the
interference contribution.

At lowest order, each parton is identified with a jet and the photon is
automatically isolated from the quark jet by requiring a non-zero
transverse momentum of the photon or jet in the $\gamma^{\ast} p$
center-of-mass frame. Therefore the photon fragmentation is absent in
this order. 

\subsection{Subprocesses to Next-to-Leading Order}

\begin{figure}[bh] 
\unitlength 1mm
\begin{picture}(160,40)
\put(31,-1){\epsfig{file=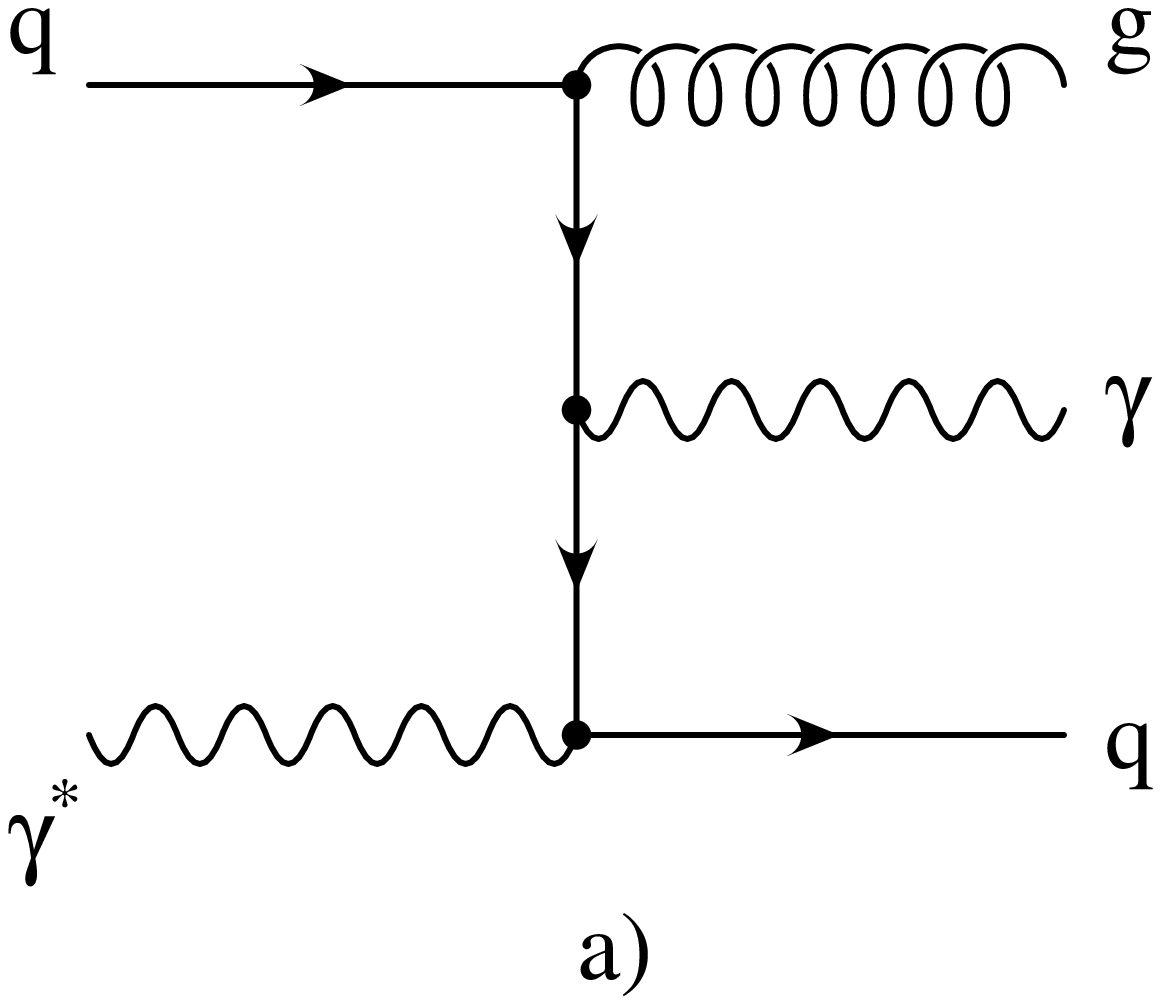,width=4cm}}
\put(91,-1){\epsfig{file=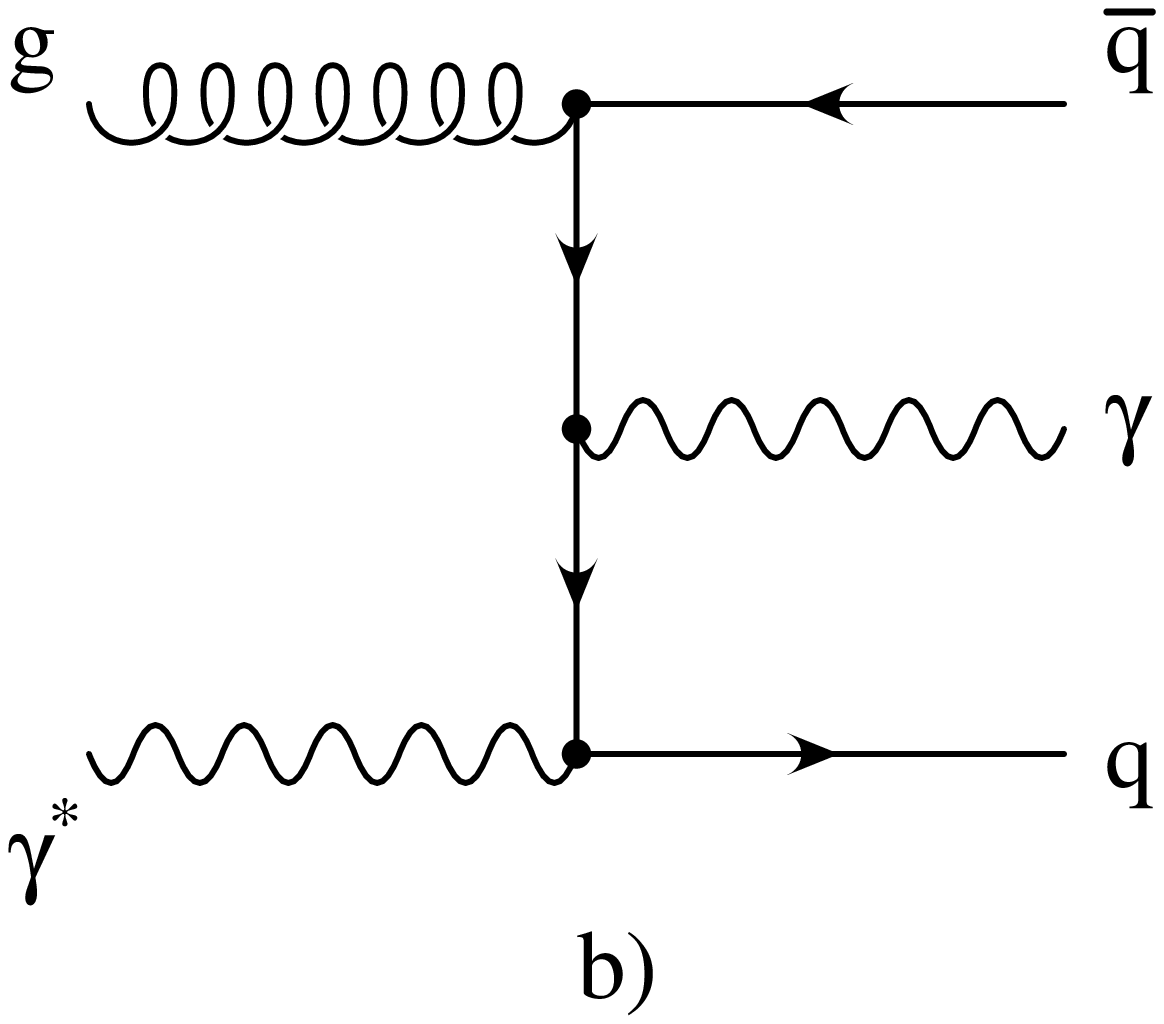,width=4cm}}
\end{picture}
\caption{Examples of Feynman diagrams for higher-order processes:
  a) $\gamma^{\ast} + q \rightarrow q + g + \gamma$,  
  b) $\gamma^{\ast} + g \rightarrow q + \bar{q} + \gamma$.} 
\label{fig2}
\end{figure}

At next-to-leading order, processes with an additional gluon, either
emitted into the final state or as incoming parton, have to be taken
into account:
\begin{equation}
e(p_1) + q(p_3) \rightarrow e(p_2) + q(p_4) + \gamma(p_5) + g(p_6),
\label{hoprocess1}
\end{equation}
\begin{equation}
e(p_1) + g(p_3) \rightarrow e(p_2) + q(p_4) + \gamma(p_5) + \bar{q}(p_6),
\label{hoprocess2}
\end{equation}
where the momenta of the particles are again given in parentheses.
Examples of diagrams for $\gamma^{\ast} q \rightarrow q\gamma g$ and
$\gamma^{\ast} g \rightarrow q\gamma\bar{q}$ are shown in Fig.\ 
\ref{fig2}. In addition, virtual corrections (one-loop diagrams at
$O(\alpha_s)$) to the LO processes (\ref{loprocess}) have to be
included. The complete matrix elements for the processes
(\ref{hoprocess1}) and (\ref{hoprocess2}) are given in \cite{dmich}. The
processes (\ref{hoprocess1}) and (\ref{hoprocess2}) contribute both to
the $\gamma + (1+1)$-jet cross section, as well as to the cross section
for $\gamma + (2+1)$-jets in the final state. In the latter case each
parton in the final state builds a jet on its own, whereas for $\gamma +
(1+1)$-jets a pair of final state partons is experimentally unresolved.
The criteria for combining two partons into one jet will be introduced
later. The contributions (\ref{hoprocess1}) and (\ref{hoprocess2}) as
well as the virtual corrections to (\ref{loprocess}) are of order
$O(\alpha^2 \alpha_s)$.

\subsection{Fragmentation Contributions}

\begin{figure}[bh] 
\unitlength 1mm
\begin{picture}(160,40)
\put(31,-1){\epsfig{file=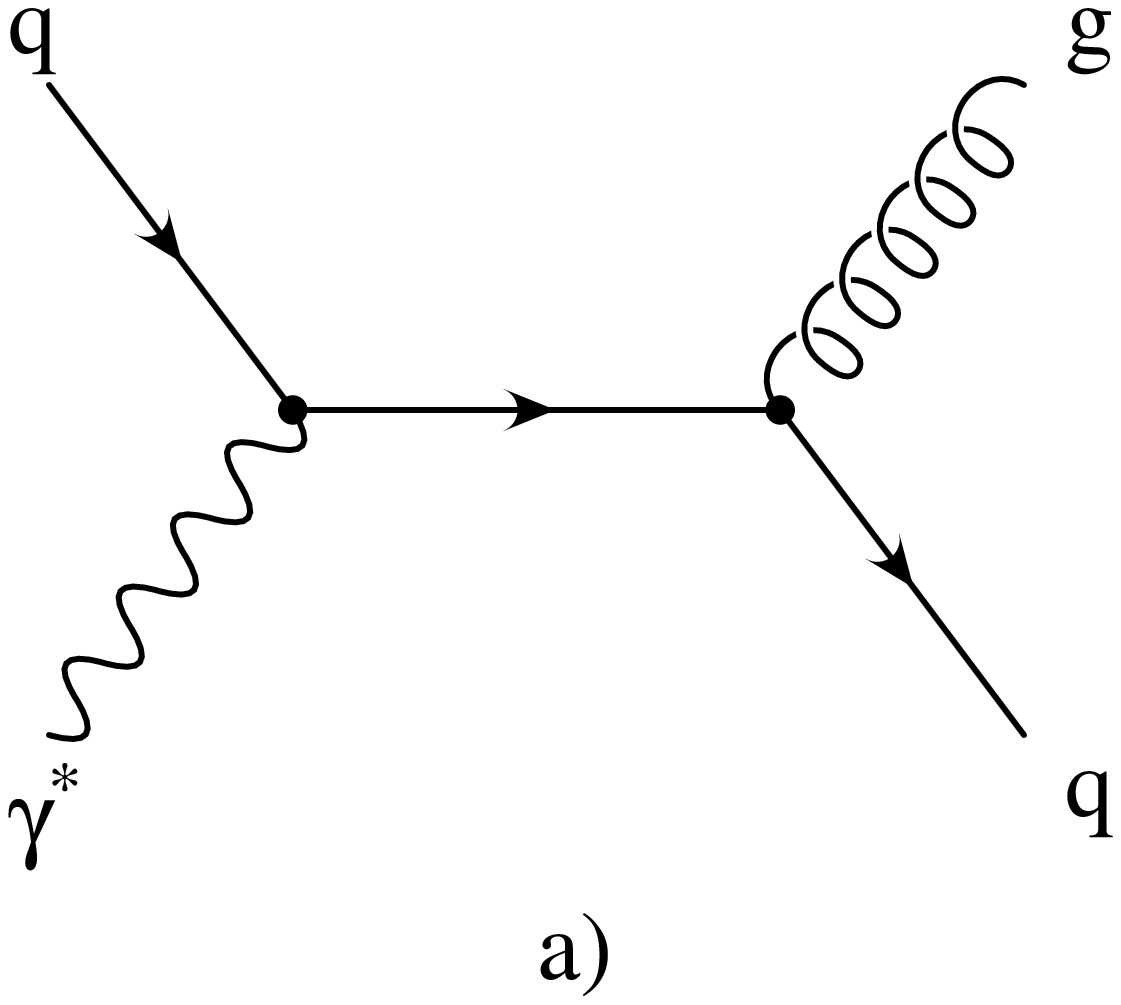,width=4cm}}
\put(91,-1){\epsfig{file=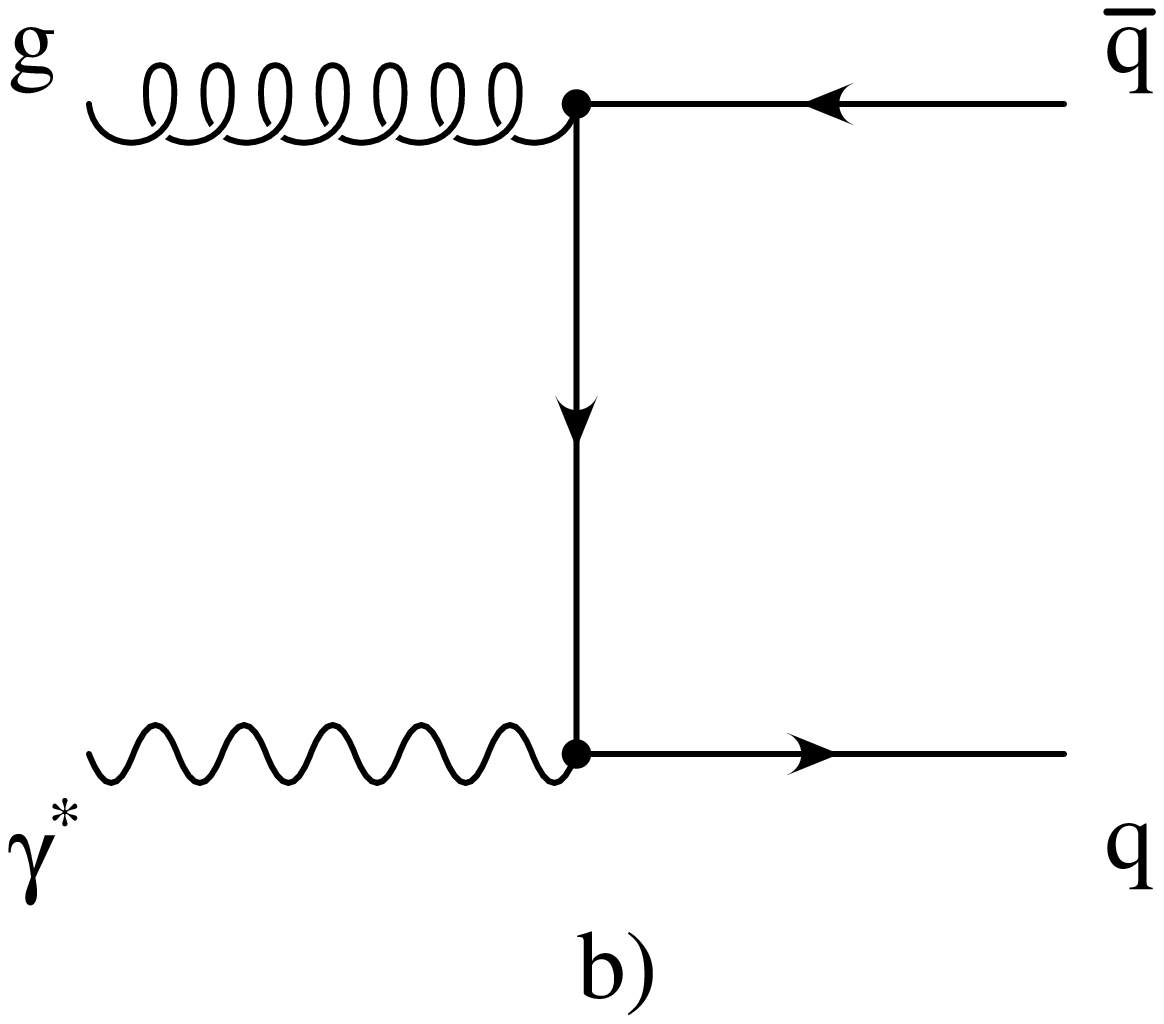,width=4cm}}
\end{picture}
\caption{Feynman diagrams giving rise to fragmentation contributions: 
  a) $\gamma^{\ast} + q \rightarrow q + g$,  
  b) $\gamma^{\ast} + g \rightarrow q + \bar{q}$.} 
\label{fig3}
\end{figure}

In addition to the direct production described in the last two
subsections, photons can also be produced through the fragmentation of a
hadronic jet into a single photon carrying a large fraction of the jet
energy \cite{koller}. This long-distance process is described in terms
of quark-to-photon and gluon-to-photon fragmentation functions which
absorb collinear singularities present in the NLO direct contributions
of section 2.2. The corresponding fragmentation processes (see Fig.\ 
\ref{fig3}) are
\begin{equation}
e(p_1) + q(p_3) \rightarrow e(p_2) + q(p_4) + g(p_6),
\label{fprocess1}
\end{equation}
\begin{equation}
e(p_1) + g(p_3) \rightarrow e(p_2) + q(p_4) + \bar{q}(p_6).
\label{fprocess2}
\end{equation}
These processes are of order $\alpha \alpha_s$ whereas the photon FF is
formally of order $\alpha$, so that the LO fragmentation contribution is
formally of order $O(\alpha^2 \alpha_s)$, i.e.\ of the same order as the
NLO direct contribution. The fragmentation photons, sometimes also
called bremsstrahlung photons, are emitted predominantly along the
direction of motion of the parent quark or gluon. Because of the
pointlike nature of the photon-quark interaction, it is possible to
calculate the leading-logarithmic behaviour of the photon FF, including
the corrections due to additional gluon emissions. The resulting FF's
are in fact of order $O(\alpha / \alpha_s)$ since they possess a
logarithmic growth coming from the integration over the momenta of
unobserved partons.  Therefore in the leading-logarithmic approximation
the fragmentation contribution is obtained from $O(\alpha / \alpha_s)$
FF's convoluted with corresponding $O(\alpha \alpha_s)$ cross sections
for the two-body subprocesses (\ref{fprocess1}) and (\ref{fprocess2}).
The resulting contribution is thus of order $O(\alpha^2)$, i.e.\ of the
same order as the LO non-fragmentation process (\ref{loprocess}).  For
this reason it is sometimes argued that the fragmentation contribution
should be combined with the LO direct process to provide the full LO
physical cross section. Consequently, the calculation of the full cross
section up to NLO would then also require the computation of the NLO
corrections to the fragmentation contributions. On the other hand, it is
well-known, that the fragmentation contribution in LO depends strongly
on the factorization scale $\mu_F$ which, however, is cancelled to a
large extent by the $\mu_F$-dependence of the NLO contribution to the
non-fragmentation part. For this reason and also since for an isolated
photon the fragmentation contribution is small we shall take it into
account only in LO in the same way as in our previous work
\cite{GKS,GKS2}. 

The signature of the fragmentation contribution in LO is a photon
balanced by a jet on the opposite side of the event and accompanied by
nearly collinear hadrons on the same side of the event.  This means that
this contribution has a similar event structure as the LO direct
contribution.

\subsection{Calculational Details}

The calculation of the NLO corrections was performed with the help of
the phase space slicing method using a slicing parameter defined in
terms of invariant masses. With this method it is straightforward to
introduce the photon isolation requirement as well as to implement a 
jet definition which separates $\gamma + (1+1)$-jet from $\gamma +
(2+1)$-jet final states. Phase space slicing based on invariant masses
is also used to separate the photon-quark collinearly singular regions,
however, using another independent cutoff parameter.  The technical
steps to apply the phase space slicing method in the present case are
described in the following.

In the calculation of the contribution to the $\gamma + (1+1)$-jet cross
section we encounter the well-known infrared singularities. They appear
in those phase space regions where two partons are degenerate to one
parton, i.e.\ when the gluon becomes soft or two partons become
collinear to each other. The singularities are assigned either to the
initial state (ISR) or to the final state (FSR).  Contributions
involving the product of an ISR and a FSR factor are separated by
partial fractioning. The FSR singularities cancel against singularities
from virtual corrections to the LO process (\ref{loprocess}). For the
ISR singularities, the cancellation is incomplete and the remaining
singular contributions have to be factorized and absorbed into the
renormalized PDF's of the proton.

To carry out these steps, the singularities are isolated in an analytic
calculation using dimensional regularization. Since the corresponding
calculations are too difficult for the exact cross sections of the
processes (\ref{hoprocess1}) and (\ref{hoprocess2}) an approximate
solution is required. To achieve this, the phase space slicing
\cite{slicing} is used first to separate the singular regions in the
4-particle phase space. Then, in these regions the matrix elements are
approximated by their most singular contributions. Only for these
approximate expressions and only in the singular regions the calculation
is performed analytically. The separation of singular regions is
obtained by applying a slicing cut $y_0^J$ to the scaled invariant
masses $y_{ij}$, where $y_{ij} = (p_i + p_j)^2 / W^2_{\rm had}$ and the
mass of the hadronic final state $W_{\rm had}$ is defined by $W_{\rm
  had}^2 = (P + q - p_5)^2$. The cut $y_0^J$ must be chosen small enough
so that terms of order $O(y_0^J)$ which are neglected in the singular
approximation are so small that an accuracy of a few per cent is
achievable for the final result.  Outside the singular regions the
integrations are done numerically without any approximation and with 4
space-time dimensions.  Physical cross sections, as defined in the next
section, are obtained by adding the contributions from singular and
non-singular regions as well as the virtual contributions and
subtracting the remaining ISR collinear singularities. In the final
results, the dependence on the slicing parameter $y_0^J$ cancels.  This
means the cut-off $y_0^J$ is purely technical. The independence on the
slicing cut $y_0^J$ has been checked by explicit calculation for some
special photon plus jet cross sections in \cite{KMS}.  Further details
and the derivation of the two-body matrix elements in the singular
region together with the cancellation of the soft and collinear poles
can be found in \cite{dmich,KMS}.

In addition, the squared matrix elements for the processes
(\ref{hoprocess1}) and (\ref{hoprocess2}) have photonic infrared and
collinear singularities, i.e.\ singularities due to soft or collinear
photons. Since we require the photon to be observed in the detector the
infrared singularity can not occur. In the numerical analysis we will
introduce this condition by requiring a minimum on the transverse
momentum of the photon. This cut removes also all collinear
singularities due to initial state radiation.
 
Final state collinear singularities due to photons are present and are
treated again with the help of the phase space slicing method in a
similar way as the quark-gluon collinear contributions. The phase space
slicing parameter used to treat the photonic singularities can be chosen
independently and is denoted by $y_0^{\gamma}$. As before, it has to be
chosen very small so that the matrix element can be approximated by its
singular part. For the subprocess (\ref{hoprocess1}), the phase space
slicing is described by the squared invariant masses $y_{45} = (p_4 +
p_5)^2/W^2_{\rm had}$ . In the gluon-initiated process
(\ref{hoprocess2}) one has two singular regions which are controlled by
the variables $y_{45}$ and $y_{56}$, respectively. In the regions
$y_{45}$, $y_{56} > y_0^{\gamma}$ the cross section is evaluated
numerically in the same way as in \cite{KMS} where these cuts were
introduced as physical isolation cuts on the photon with sufficiently
large isolation parameter $y_0^{\gamma}$. In this work the cuts on
$y_{45}$ and $y_{56}$ are only technical since we include the
contribution to the matrix element also in the regions $y_{45} <
y_0^{\gamma}$ and $y_{56} < y_0^{\gamma}$. In these regions the matrix
elements are collinearly divergent. The singularities are regulated by
dimensional regularization, allowing us to absorb their divergent parts
into the bare photon FF to yield the renormalized photon FF denoted by
$D_{q \rightarrow \gamma}$. For the process (\ref{hoprocess1}) this
procedure results in a contribution of the following form:
\begin{equation}
|M|^2_{\gamma^* q \rightarrow \gamma qg} 
= |M|^2_{\gamma^* q \rightarrow qg}
  \otimes D_{q \rightarrow \gamma}(z).     
\label{m2frag}
\end{equation}
The matrix element $|M|^2$ on the right-hand side of (\ref{m2frag}) is
the matrix element for the process $\gamma^{\ast} q \rightarrow qg$
whose Feynman diagram is shown in Fig.\ \ref{fig3}a. There exists a
similar expression for the subprocess $\gamma^{\ast} g \rightarrow
\gamma q\bar{q}$ (Fig.\ \ref{fig3}b). The photon FF $D_{q \rightarrow
  \gamma}(z)$ in (\ref{m2frag}) is given by \cite{glover1,GGdR2}
\begin{equation}
D_{q \rightarrow \gamma}(z)= D_{q \to \gamma}(z,\mu_{F}^2) + 
\frac{\alpha e_q^2}{2\pi}
 \left(P^{(0)}_{q\gamma}(z)\ln\frac{z(1-z)y_0^{\gamma}W^2}{\mu_F^2} 
         + z\right)\, .
\label{dqg}  
\end{equation} 
$D_{q\rightarrow \gamma}(z,\mu_F^2)$ in (\ref{dqg}) stands for the
non-perturbative FF describing the transition $q \rightarrow \gamma$ at
the factorization scale $\mu_F$. This function will be specified in the
next section. The second term in (\ref{dqg}), if substituted in
(\ref{m2frag}), is the finite part due to the collinear photon-quark
(-antiquark) contribution to the matrix element $|M|^2_{\gamma^* q
  \rightarrow \gamma qg}$ integrated in the region $y_{45} <
y_0^{\gamma}$ after absorption of the divergent part into the
non-perturbative FF.

Again the parameter $y_0^{\gamma}$ is only a parameter used in
intermediate steps of the calculation, introduced to separate divergent
from finite contributions; the $y_0^{\gamma}$-dependence in (\ref{dqg})
is canceled by the dependence of the numerically computed $\gamma +
(1+1)$-jet cross section restricted to the region $y_{45} >
y_0^{\gamma}$. Since the corresponding contributions to the matrix
element in (\ref{m2frag}) are calculated in the collinear approximation,
the result is valid only up to terms of order $O(y_0^{\gamma})$. This
requires to choose a very small value for $y_0^{\gamma}$. In Ref.\ 
\cite{GKS2} it has been explicitly shown that the sum of all terms for
the photon plus jet cross section becomes independent of $y_0^{\gamma}$
when $y_0^{\gamma}$ is chosen small enough.

In (\ref{dqg}), $P_{q\gamma}^{(0)}$ is the LO quark-to-photon splitting
function
\begin{equation}
P_{q\gamma}^{(0)}(z) = \frac{1 + (1-z)^2}{z}
\label{pqgamma}
\end{equation}
and $e_q$ is the electric charge of quark $q$. The variable $z$ denotes
the fraction of the quark momentum carried away by the photon. If the
photon is emitted from the final state quark with 4-momentum
$p_4^{\prime} = p_4 + p_5$, then $z$ can be related to the invariants
$y_{35}$ and $y_{34}$:
\begin{equation}
 z = \frac{y_{35}}{y_{34^{\prime}}} = 
     \frac{y_{35}}{y_{34}+y_{35}} \, .
\label{zgamma}
\end{equation}
The fragmentation contribution is proportional to the cross section for
$\gamma^{\ast} q \rightarrow qg$ which is $O(\alpha\alpha_s)$ and is
well-known. It must be convoluted with the function in (\ref{dqg}) to
obtain the contribution to the cross section for $\gamma^{\ast} q
\rightarrow q\gamma g$ at $O(\alpha^2\alpha_s)$. Equivalent formulas are 
used to calculate the fragmentation contributions to the channel
(\ref{hoprocess2}) and in the case where the quarks in the initial and
final state are replaced by an antiquark in (\ref{hoprocess1}). 

\section{Results}

\subsection{Kinematical Selection Cuts and Other Input}

The results for the cross sections which will be presented in the
following subsections are obtained for energies and kinematical cuts
appropriate for HERA experiments. The energies of the incoming electron
(positron) and proton are $E_e = 27.5$ GeV and $E_P = 820$ GeV,
respectively. The cuts on the DIS variables are chosen as in our
previous works \cite{GKS,GKS2}:
\begin{equation}
\begin{array}{ll}
   Q^2 \geq 10~{\rm GeV}^2, & W_{\rm had} \geq 10~{\rm GeV}\, ,  
\nonumber \\[2ex]
   10^{-4} \leq x \leq 0.5,~~~ & 0.05 \leq y \leq 0.95 
\end{array}
\label{cuts1}
\end{equation}
where $Q^2 = -q^2$ and $q$ is the electron momentum transfer, $q = p_1 -
p_2$ as usual.

To reduce the background from leptonic radiation \cite{KMS} we require
\begin{equation}
90^{\circ} \leq \theta_{\gamma} \leq 173^{\circ},~~~~~~
\theta_{\gamma e} \geq 10^{\circ}
\label{cuts2}
\end{equation}
where $\theta_{\gamma}$ is the emission angle of the photon measured
with respect to the momentum of the incoming electron in the HERA
laboratory frame. The cut on $\theta_{\gamma e} (=\theta_{25})$, the
angle between the momenta of the photon and the scattered electron,
suppresses radiation from the final state electron. In the case of
photon plus jet cross sections, the photon and the hadron jets $J$ are
required to have minimal transverse momenta in the $\gamma^{\ast}p$
center-of-mass system (i.e.\ in the rest system of $p_1 - p_2 + P$ where 
the remnant has $p_{T,r} = 0$),
\begin{equation}
 p_{T,\gamma} \geq 5~{\rm GeV},~~~~~~~~~p_{T,J} \geq 6~{\rm
   GeV}\, . 
\label{ptcuts}
\end{equation}
Note that an event is rejected if we do not find at least one jet with
$p_{T,J}$ above the cut in (\ref{ptcuts}). If there are two partons with
$p_{T,J} \geq 6$ GeV, the event has a chance to be treated as a $\gamma
+ (2+1)$-jet event. If, after trying to recombine partons into jets (see
below), only one jet has $p_{T,J} \geq 6$ GeV, the event contributes to
the $\gamma + (1+1)$-jet class. In the latter case, it may happen that
there is an additional parton not combined into a jet with $p_{T,i} < 6$
GeV (in a JADE-like jet algorithm applied in the HERA laboratory frame,
such a low-$p_T$ parton would be recombined with the remnant jet in most
cases).

Different values of minimal $p_T$'s for the photon and the jet have to
be chosen in order to avoid the otherwise present infrared sensitivity
of the NLO predictions \cite{GKS}. This point will be studied in detail
later.  For inclusive photon cross sections, i.e.\ in the case where we
do not perform a jet analysis of the hadronic final state, we replace
the second of the conditions in (\ref{ptcuts}) by a cut on the sum of
transverse energies of all final state partons
\begin{equation}
E_{T} = \sum_{i = q,\bar{q},g} \left| p_{T,i} \right| \geq 6~{\rm GeV}
\, . 
\label{ETcuts}
\end{equation} 

The PDF's of the proton are taken from \cite{mrs} (MRST). Recent updates
of available PDF parametrizations \cite{mrs99,cteq4+5} (MRS99, CTEQ5) do
not lead to markedly different results (roughly a 2\,\% (4\,\%) increase
of the total cross section for MRS99 (CTEQ5) with respect to MRST; a
version with enhanced $d/u$ ratio of MRS99 does not lead to observable
differences in the range of $x$ and $Q^2$ considered here).  $\alpha_s$
is calculated from the two-loop formula with the same $\Lambda$-value
($\Lambda_{\overline{MS}}(n_f=4) = 300$ MeV) as used in the MRST
parametrization of the proton PDF. The scale in $\alpha_s$ and the
factorization scale are set equal to each other and fixed to the largest
$p_{T,J}$, except when we present results with other scale choices. For
completeness, we also mention that the slicing cuts have been fixed at
$y_0^J = 10^{-4}$ and $y_0^{\gamma} = 10^{-5}$.

The dependence of the $\gamma+$jet cross sections on the choice of the
photon FF has been studied earlier \cite{GKS2}. In the present work we
choose the FF of Bourhis et al.\ \cite{bfg} (BFG). This FF has been
compared to the ALEPH $e^+e^- \rightarrow \gamma + 1$-jet cross section
\cite{aleph} and also to the inclusive photon cross section measured by
OPAL \cite{opal}. Both data sets agreed well with predictions based on
the BFG parametrization \cite{opal,GGdR2}. In \cite{GKS2} we studied
the cross section differential with respect to the fraction of momentum
$z_{\gamma}$ carried by a photon inside a jet for several other photon
FF's besides the BFG parametrization.

Photon isolation is implemented with the help of the cone isolation
method similar to the one used in the ZEUS experiment for photon
production with almost real photons \cite{HERA-data}. This method
restricts the hadronic energy allowed in a cone around the jet axis of
the jet containing the photon.  The same method is used also to define
jets emerging in the event sample of $\gamma+2$-parton-level jets when
two partons are combined.  In the $\gamma^{\ast} p$ center-of-mass
frame, two partons $i$ and $j$ are combined into a jet $J$, when they
obey the cone constraints $R_{i,J} < R$ and $R_{j,J} < R$, where
\begin{equation}
 R_{i,J} =\sqrt{(\eta_i-\eta_J)^2 + (\phi_i-\phi_J)^2}\, .
\label{Rij}
\end{equation}
$\eta_J$ ($= - \ln \tan (\theta_J/2)$) and $\phi_J$ are the rapidity
(polar angle $\theta_J$) and azimuthal angle of the recombined jet which
are obtained from the formulae
\begin{equation}
\begin{array}{l}
p_{T,J} = p_{T,i} + p_{T,j}\, , \\[2ex]
\displaystyle
\eta_J = \frac{\eta_i p_{T,i} + \eta_j p_{T,j}}{p_{T,i} + p_{T,j}} \, ,
\\[2ex]  
\displaystyle
\phi_J = \frac{\phi_i p_{T,i} + \phi_j p_{T,j}}{p_{T,i} + p_{T,j}} \, . 
\end{array}
\label{jet-variables}
\end{equation}
For most of the results we choose $R=1$. If not, we shall state the
value of $R$ explicitly. The azimuthal angle is defined with respect to
the scattering plane given by the momentum of the beam and the momentum
of the scattered electron. The rapidity is always defined positive in
the direction of the proton remnant. The photon is treated like any
other parton in the recombination process, i.e.\ when in (\ref{Rij}) $i$
or $j$ is the photon, then $J$ is called the photon-jet.  To qualify a
jet as a photon-jet, we restrict the hadronic energy in the jet by
\begin{equation}
z_{\gamma} = \frac{p_{T,\gamma}}{p_{T,\gamma}+p_{T,{\rm had}}}
= 1 - \epsilon_{\rm had} \geq 1 - \epsilon_{\rm had}^0 = z_{\rm cut}.
\label{zcut}
\end{equation}
$p_{T,\gamma}$ and $p_{T,{\rm had}}$ denote the transverse momenta of
the photon and the parton producing hadrons in this jet. Defining
$p_{T,\gamma {\rm -jet}} = p_{T,\gamma} + p_{T,{\rm had}}$, (\ref{zcut})
is equivalent to the requirement
\begin{equation}
p_{T,{\rm had}} \leq \epsilon_{\rm had}^0 \ p_{T,\gamma{\rm -jet}} , 
\label{zcut2}
\end{equation}
i.e.\ the ratio of the total $p_T$ due to other particles (partons) than
the photon is required not to exceed $\epsilon_{\rm had}^0$. For our
predictions we shall choose different values for $\epsilon_{\rm had}^0 =
(1 - z_{\rm cut})$. In \cite{HERA-data} this parameter was set equal to
$\epsilon_{\rm had}^0 = 0.11$.

It is known that the cone algorithm is ambiguous for final states with
more than three particles or partons \cite{seymour}. Since we have
maximally three partons in the final state this problem is not relevant
in our case.  However, in some cases it may happen that two partons $i$
and $j$ qualify both as two individual jets $i$ and $j$, or as a
combined jet $J$. In these exceptional cases we count only the combined
jet $J$ to avoid double counting.

The cone algorithm is problematic in experimental analyses due to its
seed-finding mechanism and due to overlapping cones. These problems are
avoided with the $k_T$ jet finding algorithm \cite{kt-algorithm}. In our
theoretical calculations, the $k_T$ algorithm can be incorporated quite
easily: partons $i$ and $j$ (where one of them may be the photon) are
combined if they fulfill the condition
\begin{equation}
R_{ij} < R ~~~ {\rm with} ~~~
R_{ij} = \sqrt{(\eta_i - \eta_j)^2 + (\phi_i - \phi_j)^2}. 
\label{Rij-kt}
\end{equation}
The resulting kinematic variables of the combined jet are calculated
with the same formulae (\ref{jet-variables}) as in the cone algorithm.
The recombination condition (\ref{Rij}) is equivalent to
\begin{equation}
R_{ij} < R p_{ij} ~~~ {\rm with} ~~~ p_{ij} = \frac{p_{T,i} +
  p_{T,j}}{\max(p_{T,i}, p_{T,j})}. 
\label{Rijp}
\end{equation}
Therefore, choosing the same value for $R$, jets obtained with the cone
algorithm (\ref{Rijp}) are slightly wider than those constructed with
the $k_T$ algorithm (\ref{Rij-kt}). In order to demonstrate that our
numerical routines work also with the $k_T$ algorithm we have calculated
some representative cross sections with this jet definition as well.

\subsection{Photon plus Jet Cross Sections}

Now we shall present our numerical results. We start with various cross
sections for the $\gamma + (n+1)$-jet final state since we think that
these cross sections, although smaller than the fully inclusive photon
cross section, will be measured first due to reduced background
problems. It is clear that in NLO the final state may consist of two or
three jets where one jet is always a photon jet. The remnant jet is not
counted since it is produced with zero transverse momentum. The
three-jet sample, equivalent to $\gamma + (2+1)$-jets in the notation of
the previous sections, consists of all $\gamma + (2+1)$-parton-level
jets which do not fulfill the cone constraint $R_{i,J} < R$ with
$R_{i,J}$ given in (\ref{Rij}). The $\gamma + (1+1)$-jet sample contains
events where two partons (possibly a photon) are recombined into one jet
or one parton does not obey the cut on transverse momenta
(\ref{ptcuts}). In the following we shall sum over the two samples with
$n=1$ and $n=2$ jets. If there are two jets in an event, we order them
according to their transverse momenta and call the one with larger
$p_T$ ``jet 1'' and the one with smaller $p_T$ accordingly ``jet 2''.
Also, from now on we will use a simplified notation and denote kinematic
variables of the jet containing the photon simply by $p_{T,\gamma}$,
$\eta_{\gamma}$ and $\phi_{\gamma}$.

\begin{figure}[th] 
\unitlength 1mm
\begin{picture}(160,80)
\put(0,-1){\epsfig{file=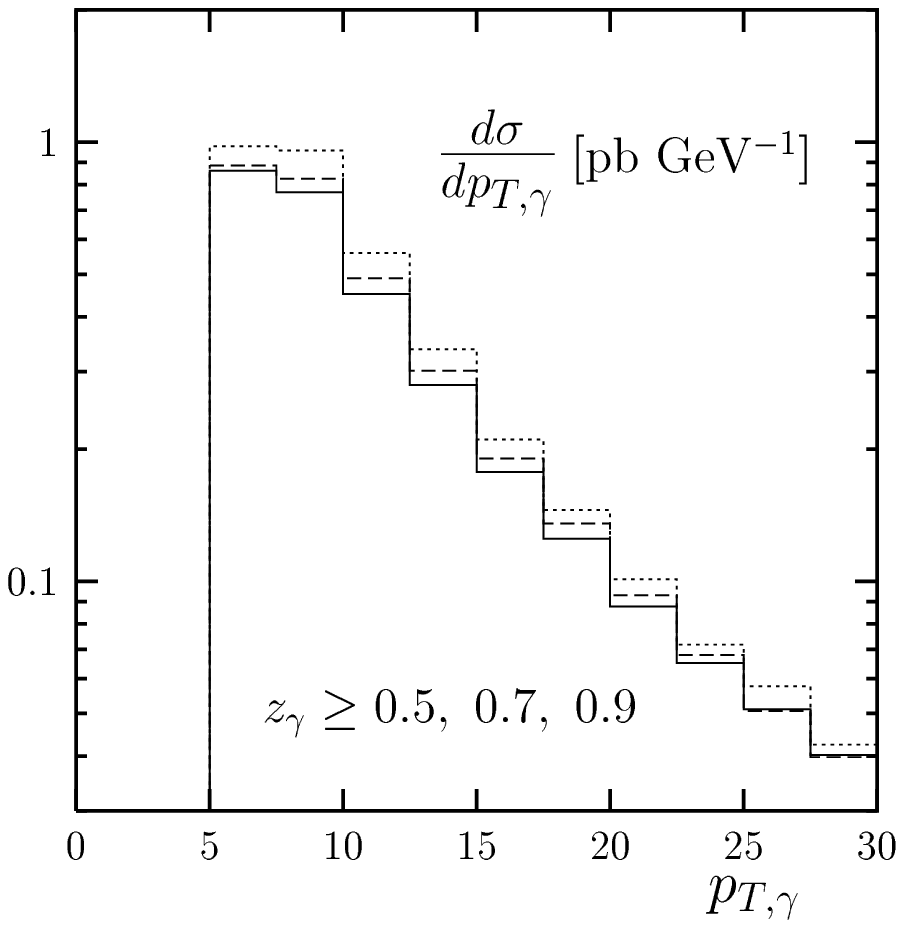,width=8cm}}
\put(40,-1){(a)}
\put(80,-1){\epsfig{file=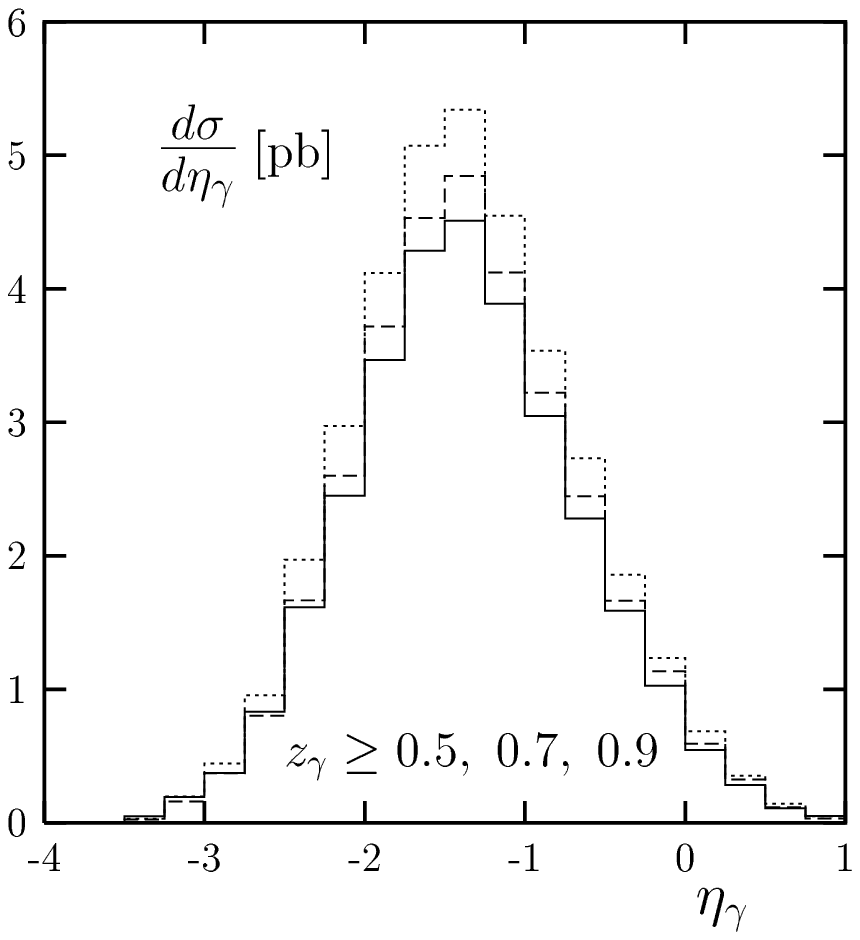,width=8cm}}
\put(120,-1){(b)}
\end{picture}
\caption{$p_T$- (a) and $\eta$- (b) distributions of the photon jet for
  $z_{\rm cut} = 0.9$, 0.7 and 0.5 (full, dashed and dotted curves).}
\label{fig4}
\end{figure}

\begin{figure}[bh] 
\unitlength 1mm
\begin{picture}(160,80)
\put(0,-1){\epsfig{file=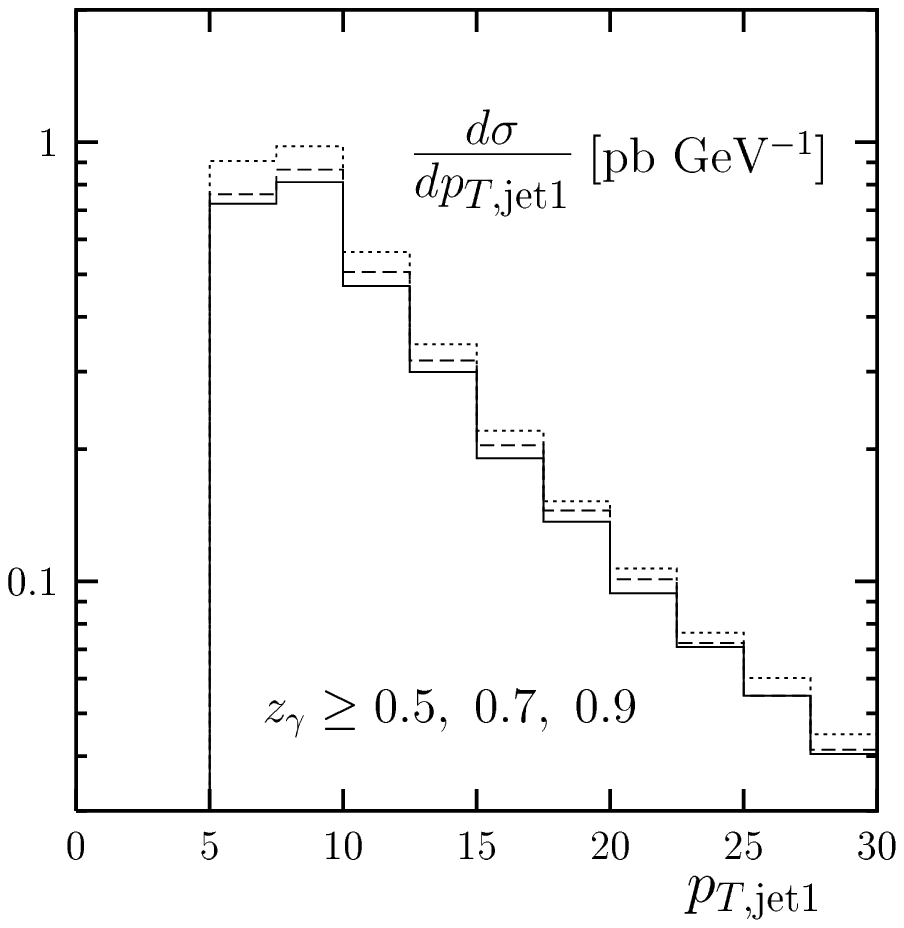,width=8cm}}
\put(40,-1){(a)}
\put(80,-1){\epsfig{file=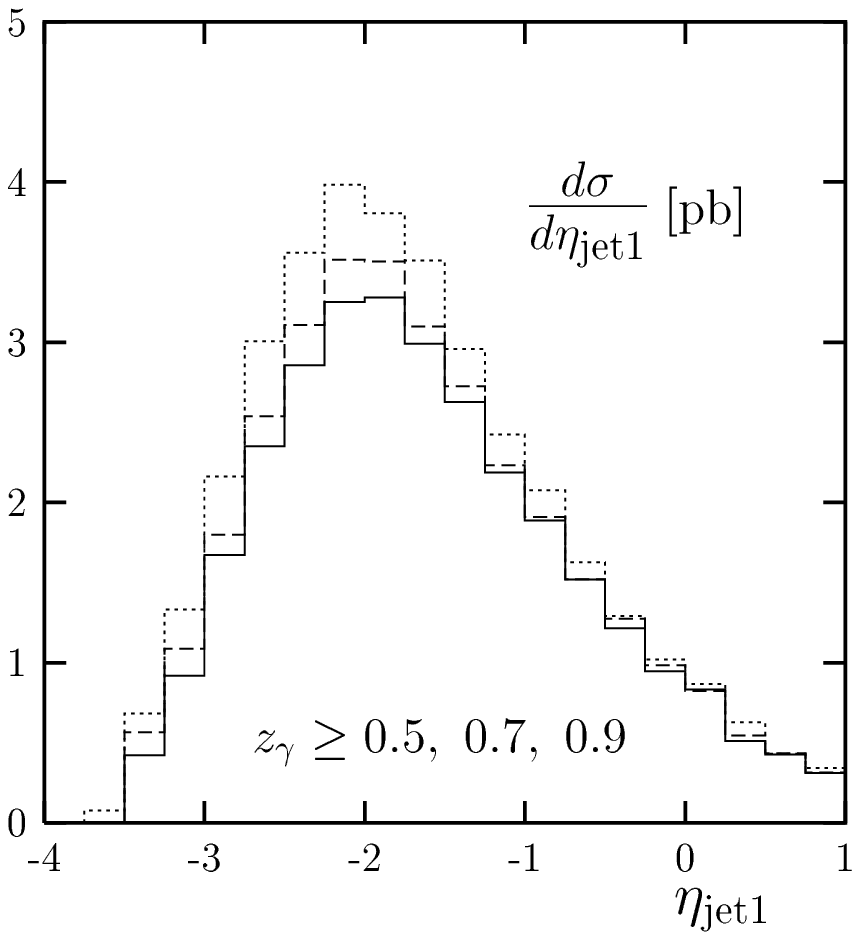,width=8cm}}
\put(120,-1){(b)}
\end{picture}
\caption{$p_T$- (a) and $\eta$- (b) distributions of the most energetic
  jet for $z_{\rm cut} = 0.9$, 0.7 and 0.5 (full, dashed and dotted
  curves).} 
\label{fig5}
\end{figure}

In Figs.\ \ref{fig4} and \ref{fig5} we show results for the $p_T$ and
$\eta$ dependence of the cross sections $d\sigma/dp_T$ and $d\sigma /
d\eta$ for the photon jet and the jet with the largest $p_T$. In each of
the four figures we have plotted three curves for three choices of
$z_{\rm cut}$ defined in (\ref{zcut}), $z_{\rm cut} = 0.5$, 0.7 and 0.9.
Together with the cone radius $R$, $z_{\rm cut}$ controls the amount of
photon isolation. As to be expected the cross section decreases with the
degree of photon isolation, i.e.\ with increasing $z_{\rm cut}$.
Specifically in Fig.\ \ref{fig4}a we present $d\sigma / dp_{T,\gamma}$
as a function of $p_{T,\gamma}$ for the three $z_{\rm cut}$ values and
for $p_{T,\gamma} \ge 5$ GeV. All other variables, in particular
$\eta_{\rm jet}$, $\eta_{\gamma}$ and $p_{T,{\rm jet}}$ are integrated
over the kinematically allowed ranges.  We see that all three cross
sections have a similar shape. In Fig.\ \ref{fig5}a $d\sigma /
dp_{T,{\rm jet}1}$ as a function of $p_{T,{\rm jet}1}$ is shown. The
qualitative behaviour of the cross sections for the different $z_{\rm
  cut}$'s is similar as in Fig.\ \ref{fig4}a.

\clearpage

\begin{figure}[tb] 
\unitlength 1mm
\begin{picture}(160,80)
\put(0,-1){\epsfig{file=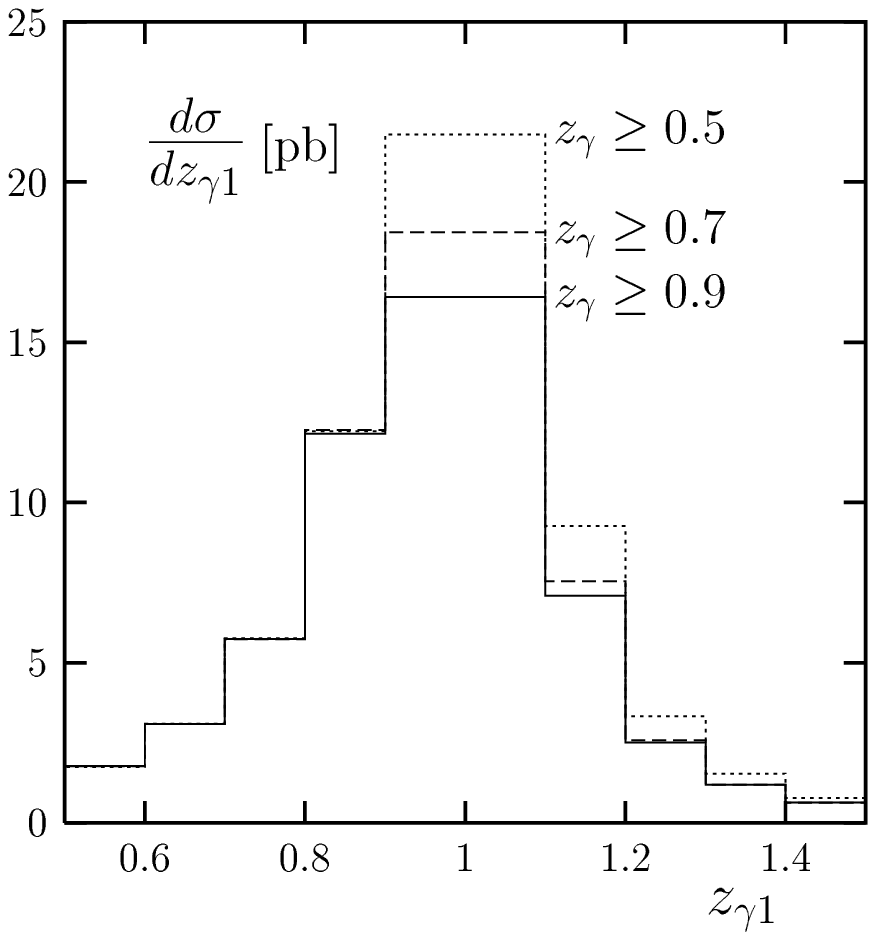,width=8cm}}
\put(40,-1){(a)}
\put(80,-1){\epsfig{file=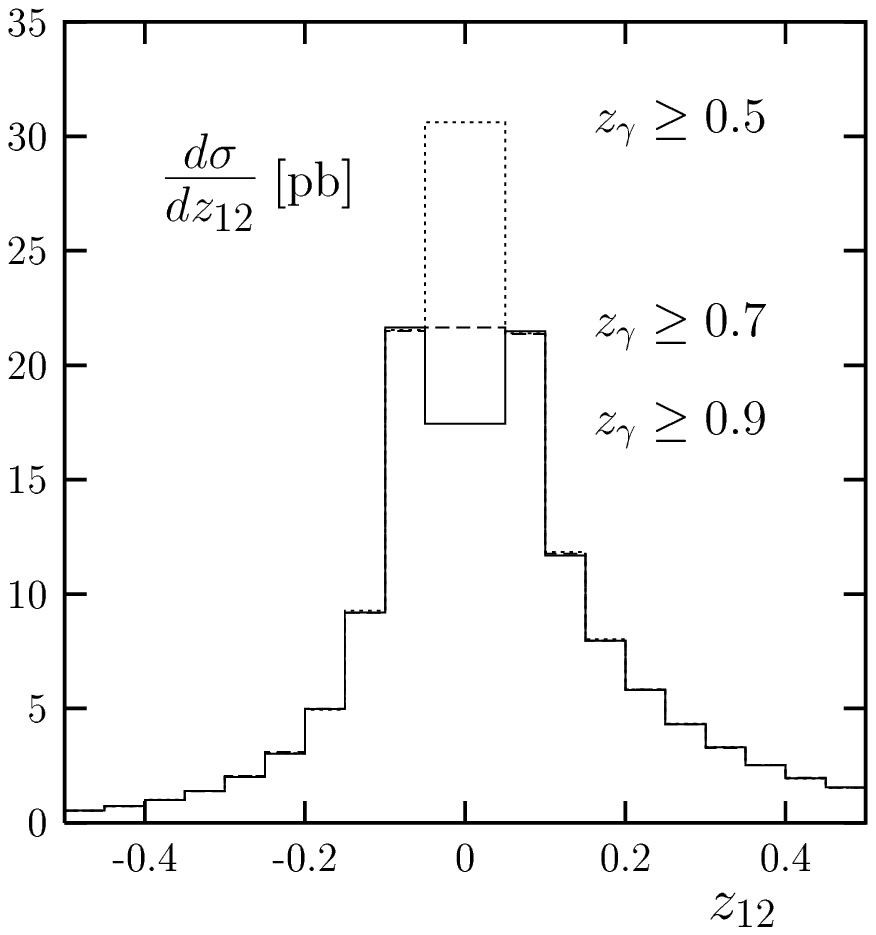,width=8cm}}
\put(120,-1){(b)}
\end{picture}
\caption{Distribution with respect to the photon-jet imbalance parameter 
  $z_{\gamma 1}$ (a) and the jet-jet imbalance parameter $z_{12}$ (b)
  for $z_{\rm cut} = 0.9$, 0.7 and 0.5 (full, dashed and dotted
  curves).}
\label{fig7}
\end{figure}

For the $\eta$ distributions, shown in Figs.\ \ref{fig4}b, \ref{fig5}b
for the photon jet and the most energetic jet, we have integrated over
$p_{T,\gamma} \ge 5$ GeV and $p_{T,{\rm jet}1} \ge 6$ GeV. The shapes of
the three curves for the different $z_{\rm cut}$ values are similar for
both the cross section $d\sigma / d\eta_{\gamma}$ in Fig.\ \ref{fig4}b
and the cross section $d\sigma / d\eta_{{\rm jet}1}$ in Fig.\ 
\ref{fig5}b.  We note that the $\eta_{{\rm jet}1}$ distribution peaks at
somewhat smaller rapidities than $d\sigma / d\eta_{\gamma}$.

Distributions with respect to $p_T$ and $\eta$ of the second jet in
$\gamma + (2+1)$-jet events do not depend on the isolation cut since in
this case each parton (photon) constitutes a jet by its own. Therefore
we show corresponding figures in the next subsection when we will
discuss the influence of the cone size $R$ on the cross sections.

In addition to predicting distributions in the transverse momenta and
the rapidities of the photon and the hadron jets we also have calculated
distributions for variables which characterize the correlation of two
jets. One of these variables is
\begin{equation}
z_{\gamma 1} = - \frac{\vec{p}_{T,\gamma}\vec{p}_{T, {\rm 
    jet}1}}{p^2_{T,{\rm jet}1}}\, .
\end{equation}
Note that $z_{\gamma 1}$ is defined with the help of the transverse
momentum of the photon jet, i.e.\ $p_{T,\gamma}$ may include a
contribution from accompanying hadronic energy. The dependence of the
cross section on $z_{\gamma 1}$ characterizes the imbalance in
transverse momentum of the photon and the most energetic jet. Similar
variables have been used before for studies in the case of photon plus
charm jet final states in $p\bar{p}$ collisions \cite{bailey} and of
two-jet production in $ep$ scattering \cite{kpoet}. The result for
$d\sigma / dz_{\gamma 1}$ is shown in Fig.\ \ref{fig7}a for three photon
isolation cuts $z_{\rm cut} = 0.5$, 0.7 and 0.9. The cuts on transverse
momenta and the cone parameters are as defined before.

For two-body processes such as the LO Compton subprocess $\gamma^{\ast}
q \rightarrow \gamma q$, the final photon and the jet have balancing
transverse momenta and the distribution is a $\delta$-function in
$(1-z_{\gamma 1})$. Also the fragmentation process contributes only to
$z_{\gamma 1} = 1$ since in this case the transverse momentum of
accompanying hadronic energy is collinear with the photon, resulting in
$p_{T,\gamma} = p_{T,{\rm bare}~\gamma} + p_{T,{\rm had}} = p_{T,{\rm
    jet}1}$.  Contributions with $z_{\gamma 1} \ne 1$ are due to the
higher-order three-body contributions. Events with $z_{\gamma 1} < 1$
typically result from configurations where a single photon is opposite
in transverse momentum to a jet consisting of two partons. Since
according to the recombination prescription (\ref{jet-variables}) the
scalar sum of transverse momenta is ascribed to the jet, not the
vectorial sum, one finds always $p_{T,\gamma} < p_{T,{\rm jet}1}$ and
thus $z_{\gamma 1} < 1$ in this case. Moreover, the photon is never
accompanied by a hadronic parton in this case and events with $z_{\gamma
  1} < 1$ are consequently not affected by the isolation cut. On the
other hand, events with $z_{\gamma 1} > 1$ are predominantly due to
configurations with a photon-jet consisting of a photon and a quark or
gluon opposite to a jet consisting of a single parton. In this case, a
variation of the photon isolation cut $z_{\rm cut}$ has a strong effect
on the differential cross section.

These features are clearly visible in Fig.\ \ref{fig7}a.  The cross
section increases when lowering $z_{\rm cut}$, i.e.\ when larger
fragmentation contributions are included. We notice that the $z_{\gamma
  1}$ distribution is not symmetric around $z_{\gamma 1} = 1$. The cross
section for $z_{\gamma 1} < 1$ is larger than for $z_{\gamma 1} > 1$,
becoming more and more symmetric for less restrictive isolation cuts on
$z_{\gamma}$. The residual asymmetric behaviour of this distribution for
vanishing isolation cut is a dynamical property of the underlying cross
section.

In the region of $z_{\gamma 1}$ near unity, one of the two final state
partons of three-body contributions (not the photon) becomes soft and
thus this region is sensitive to soft-gluon effects. In our calculation
with an invariant mass cut slicing parameter $y_0^J$ these soft-gluon
corrections are considered as two-body contributions. Their contribution
depends on the slicing parameter $y_0^J$. To remove this dependence,
i.e.\ to remove the infrared sensitivity, we must include a sufficiently
large fraction of the three-body contribution from outside $z_{\gamma 1}
= 1$.  Therefore we averaged the $z_{\gamma 1}$ distribution over
sufficiently large bins and studied the sum of the $\gamma + (1+1)$- and
$\gamma + (2+1)$-jet cross sections $d\sigma / dz_{\gamma 1}$. We have
chosen a bin width of $\Delta z_{\gamma 1} = 0.2$ around $z_{\gamma 1} =
1$ and $\Delta z_{\gamma 1} = 0.1$ elsewhere.

It is clear that the cross section outside the bin at $z_{\gamma 1} = 1$
has a stronger scale dependence than inside this bin since only
three-parton terms contribute. The cross section inside the bin at
$z_{\gamma 1}=1$ is a genuine NLO prediction with expected reduced scale
dependence. The scale dependence will be studied later for some other
distributions.  The $\delta$-function behaviour at LO is of course in
reality modified not only by NLO corrections, but also by
non-perturbative effects originating from hadronization and a possible
intrinsic transverse momentum of the initial parton. Our calculation
does not include these latter effects.

In Fig.\ \ref{fig7}b we show the cross section $d\sigma / dz_{12}$ where
the variable
\begin{equation}
z_{12} = - \frac{\vec{p}_{T,{\rm jet}1} \vec{p}_{T,{\rm jet}2}}%
                {p^2_{T,{\rm jet}1}}
\label{z12}
\end{equation}
with $p_{T,{\rm jet}1} > p_{T,{\rm jet}2}$ measures the correlation
between the two jets in $\gamma + (2+1)$-jet events. This cross section
peaks at $z_{12} = 0$ as to be expected and decreases away from $z_{12}
= 0$. The point $z_{12} = 0$ is the point with no second jet, i.e.\ the
pure $\gamma + (1+1)$-jet region. This region is again infrared
sensitive. Therefore we integrated here over the bin $-0.05 < z_{12} <
0.05$. Outside this region we chose a bin size of $\Delta z_{12} =
0.05$. Note that the distribution shown in Fig.\ \ref{fig7}b includes
the contribution from low-$p_T$ partons with $p_T$ below the cut in
(\ref{ptcuts}). We have also calculated the $z_{12}$-distribution
restricting to $\gamma + (2+1)$-jet events where both jets have $p_T \ge
6$ GeV. In this case, the distribution would extend to larger values of
$z_{12}$ with a maximum at $z_{12} \simeq 0.6$. The asymmetric behaviour
of the curve visible in Fig.\ \ref{fig7}b, i.e.\ the tail at large
$z_{12}$, is due to the contribution from these $\gamma + (2+1)$-jet
events.
\begin{figure}[tb] 
  \unitlength 1mm
\begin{picture}(160,78)
\put(40,-3){\epsfig{file=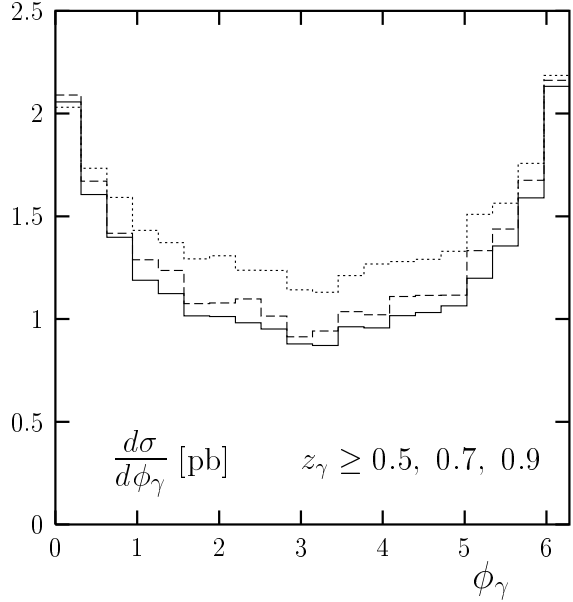,width=8cm}}
\end{picture}
\caption{$\phi$-distribution of the photon jet for $z_{\rm cut} = 0.9$,
  0.7 and 0.5 (full, dashed and dotted curves).}
\label{fig8}
\vspace*{-0.5cm}
\end{figure}

Another interesting variable might be the azimuthal angle
$\phi_{\gamma}$ of the emitted photon. We define $\phi_{\gamma}$ with
respect to the plane spanned by the momenta of the beam and of the
scattered lepton. In Fig.\ \ref{fig8} we show the dependence of the
cross section on $\phi_{\gamma}$, again for the three $z_{\rm cut}$
values 0.5, 0.7 and 0.9. As is seen in this figure, the photon is
emitted dominantly at $\phi_{\gamma} = 0$. We note that the distribution
becomes flatter with decreasing $z_{\rm cut}$. It is symmetric within
the statistical accuracy of the calculation. We do not present here a
similar plot for $d\sigma / d\phi_{{\rm jet}1}$, the cross section with
respect of the azimuthal angle of the most energetic jet. It would show
a distribution which peaks at $\phi_{{\rm jet}1} = \pi$, since the
dominant contribution to the cross section comes from configurations
in which the photon and the jet with the largest $p_T$ are emitted
back-to-back.  For jet 2 there is no such correlation. The cross section
$d\sigma / d\phi_{{\rm jet}2}$ is independent of $\phi_{{\rm jet}2}$
and, in NLO, does not change with $z_{\rm cut}$. In a similar way one
can discuss the cross section as a function of $\phi_{\gamma 1} =
\phi_{\gamma} - \phi_{{\rm jet}1}$ or $\phi_{\gamma 2} = \phi_{\gamma} -
\phi_{{\rm jet}2}$.  $d\sigma / d\phi_{\gamma 1}$ is strongly peaking at
$\phi_{\gamma 1} = \pi$. Here it is again necessary to calculate
$d\sigma /d\phi_{\gamma 1}$ with a sufficiently large bin size around
$\phi_{\gamma 1} = \pi$ in order to avoid any infrared sensitivity. On
the other hand, the distribution $d\sigma / d\phi_{\gamma 2}$ is flat
around $\phi_{\gamma 2} = 0$ and decreases towards $\phi_{\gamma 2} =
\pi$.

\subsection{Cone Size Dependence of Jet Cross Sections}

So far we presented results only for the cone jet algorithm with the
cone radius fixed to $R = 1$ for both the photon jet and purely hadronic
jets. Sometimes it is advantageous to use smaller cone radii to suppress
background processes.  On the other hand the dependence of the cross
sections on the cone radius is a genuine NLO effect since LO cross
sections do not depend on the jet definition. To present an overview of
the cone radius dependence we have recalculated some of the cross
sections shown so far for $R = 1$ for two smaller radii $R = 0.5$ and
0.7. For the photon isolation parameter we fix now $z_{\rm cut} = 0.9$.

\begin{figure}[tb] 
\unitlength 1mm
\begin{picture}(160,80)
\put(0,-1){\epsfig{file=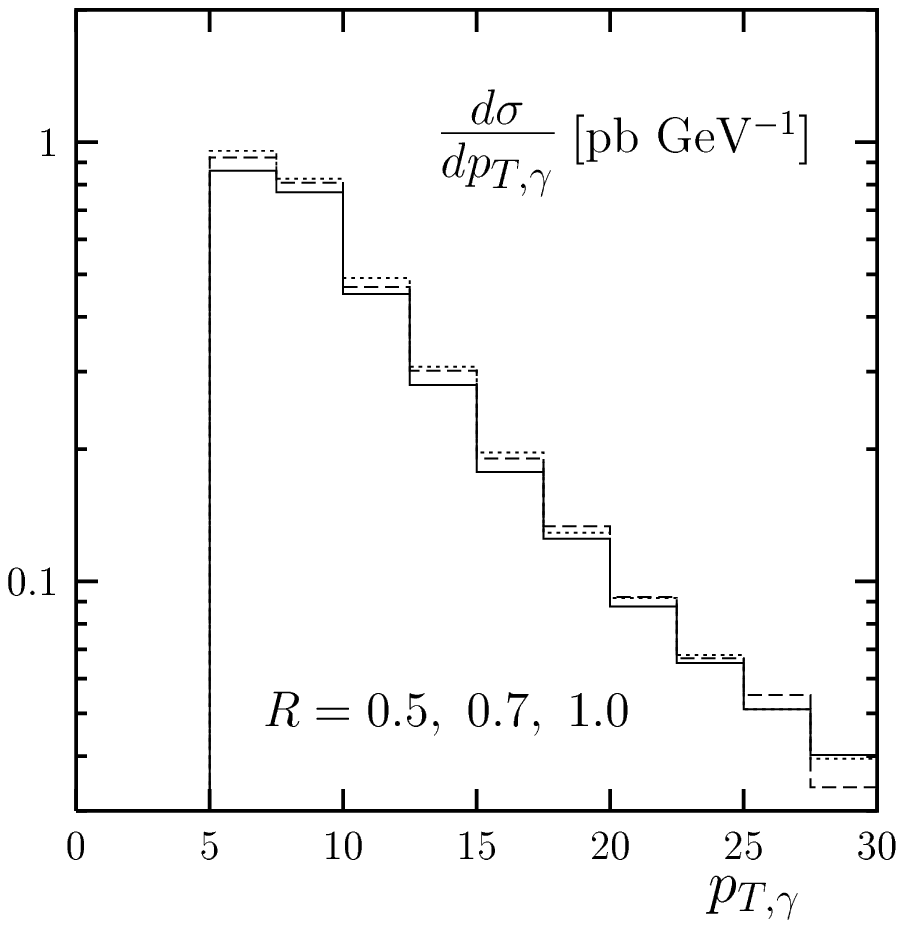,width=8cm}}
\put(40,-1){(a)}
\put(80,-1){\epsfig{file=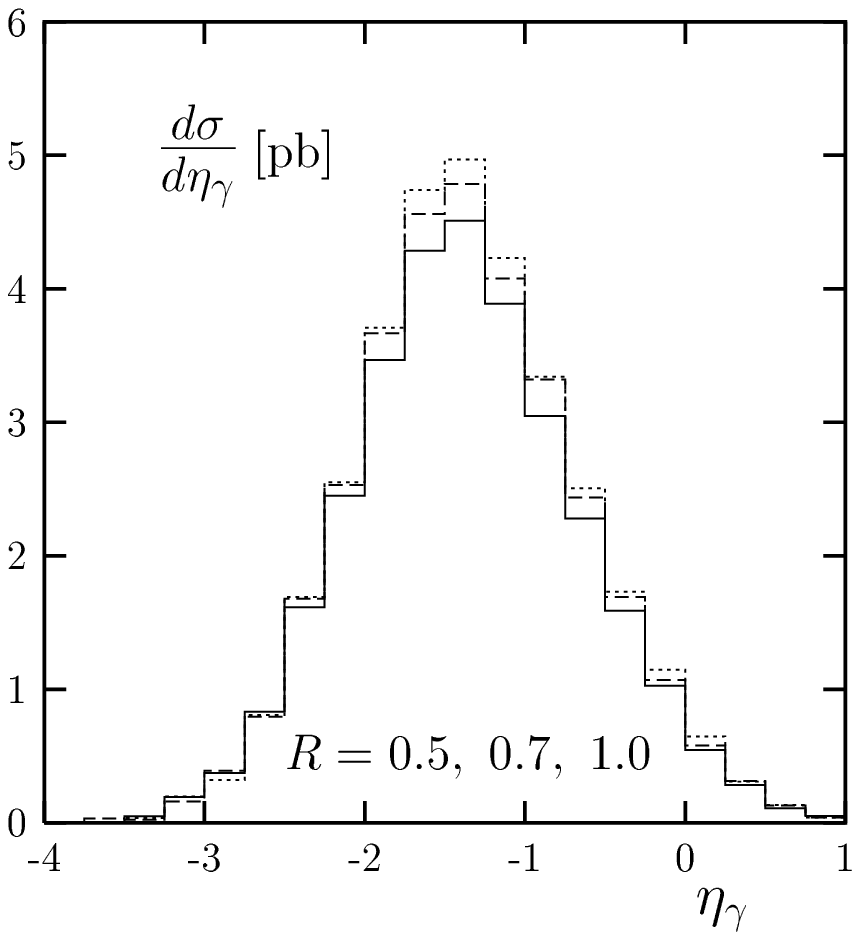,width=8cm}}
\put(120,-1){(b)}
\end{picture}
\caption{$p_T$- (a) and $\eta$- (b) distributions of the photon jet for
  $R = 1.0$, 0.7 and 0.5 (full, dashed and dotted curves).}
\label{fig10}
\end{figure}

\begin{figure}[tb] 
\unitlength 1mm
\begin{picture}(160,80)
\put(0,-1){\epsfig{file=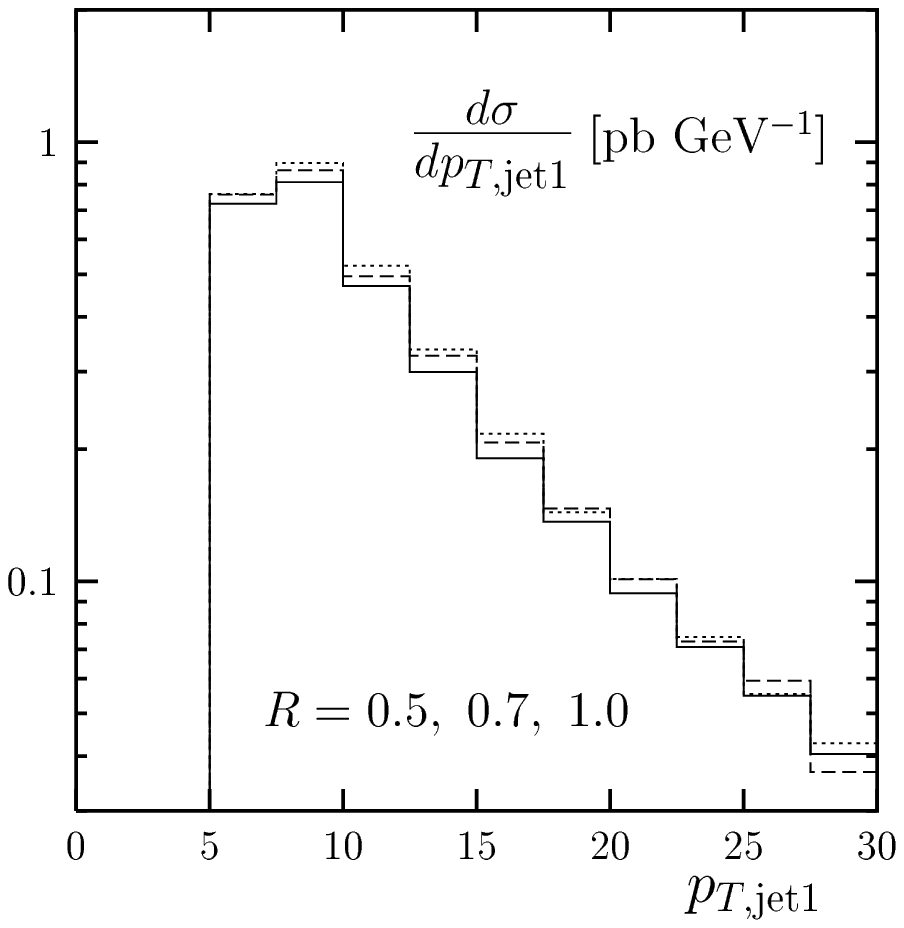,width=8cm}}
\put(40,-1){(a)}
\put(80,-1){\epsfig{file=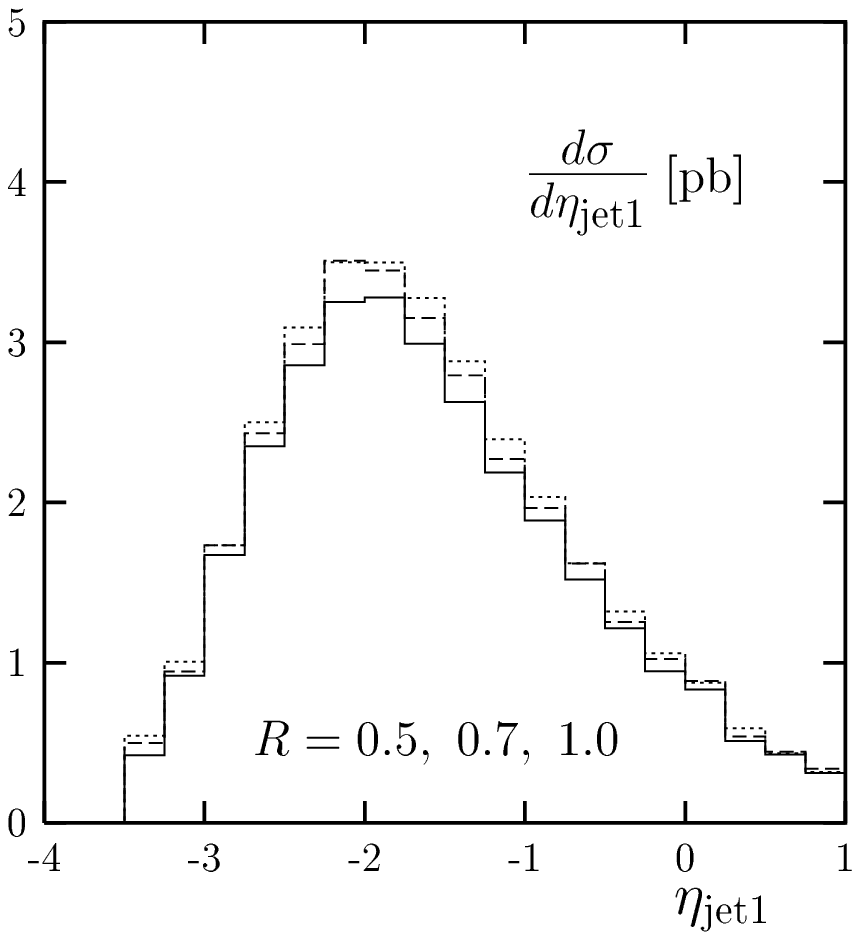,width=8cm}}
\put(120,-1){(b)}
\end{picture}
\caption{$p_T$- (a) and $\eta$- (b) distributions of the most energetic
  jet for $R = 1.0$, 0.7 and 0.5 (full, dashed and dotted curves).}
\label{fig11}
\end{figure}

\begin{figure}[tb] 
\unitlength 1mm
\begin{picture}(160,80)
\put(0,-1){\epsfig{file=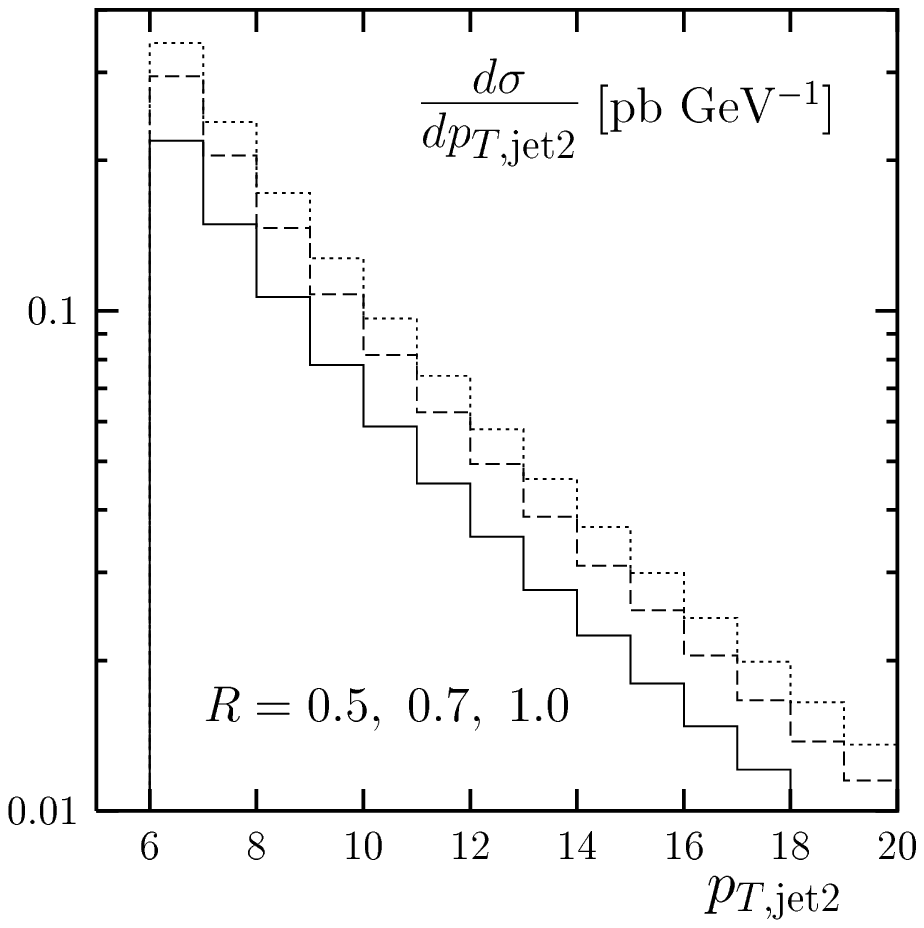,width=8cm}}
\put(40,-1){(a)}
\put(80,-1){\epsfig{file=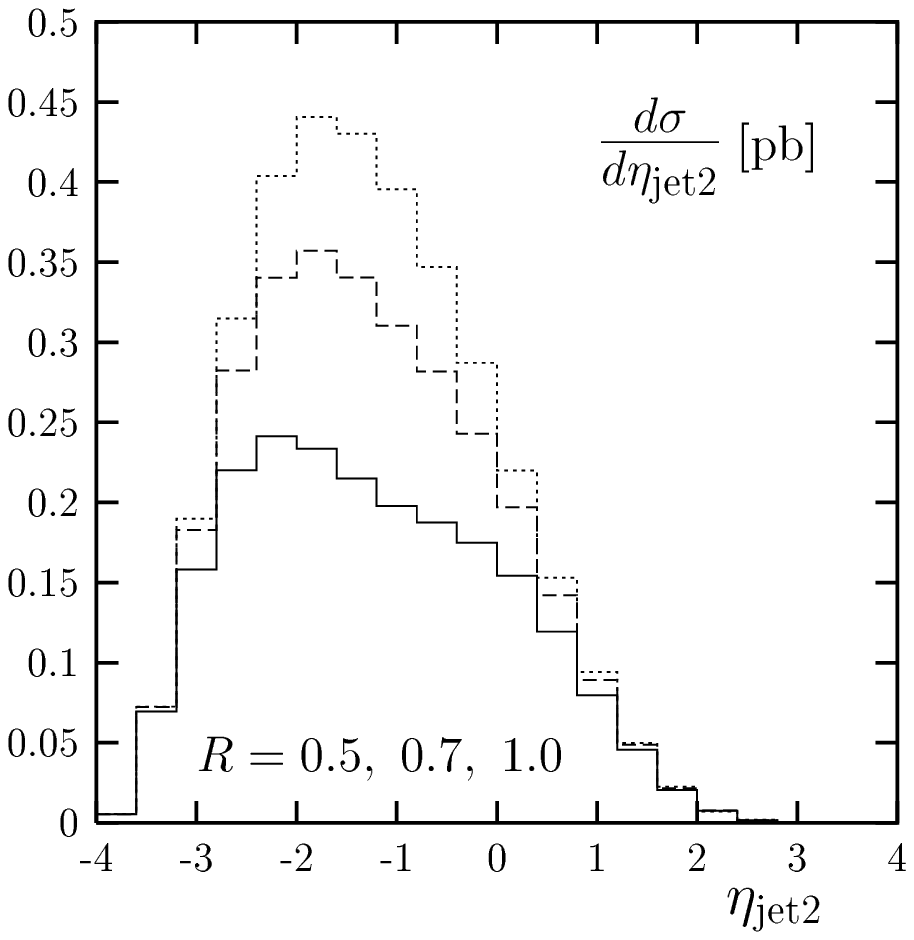,width=8cm}}
\put(120,-1){(b)}
\end{picture}
\caption{$p_T$ (a) and $\eta$ (b) distributions of the second, less
  energetic jet for $R = 1.0$, 0.7 and 0.5 (full, dashed and dotted
  curves).}
\label{fig12}
\end{figure}

In Figs.\ \ref{fig10}a, b we show the results for $d\sigma /
dp_{T,\gamma}$ and $d\sigma / d\eta_{\gamma}$ with $R = 0.5$, 0.7, and
1.0. All other cuts are chosen as before. The case with $R = 1$ and
$z_{\rm cut} = 0.9$ was shown in Figs.\ \ref{fig4}a, b already. The
distributions for the jet with largest $p_T$ are exhibited in Figs.\ 
\ref{fig11}a and b. $d\sigma / dp_{T,\gamma}$, $d\sigma /
d\eta_{\gamma}$, $d\sigma / dp_{T,{\rm jet}1}$ and $d\sigma /
d\eta_{{\rm jet}1}$ show very little dependence on the cone size $R$.
For a less restrictive isolation cut of $z_{\gamma} \ge 0.5$ these
distributions would decrease with decreasing cone radius $R$ as it is
known from other jet calculations.

In our previous work \cite{GKS} we studied the equivalent cross sections
also for the two event classes with $\gamma + (1+1)$-jets and $\gamma +
(2+1)$-jets separately. It turned out that the contribution for $\gamma
+ (1+1)$-jets is the dominant one.  This is expected, since the
contribution with $\gamma + (2+1)$-jets is an $O(\alpha_S)$ effect.
Here we present now also results for the $p_T$ and $\eta$ distributions
of jet 2 in this latter process.  The results are shown in Fig.\ 
\ref{fig12}a ($d\sigma / dp_{T,{\rm jet}2}$) and Fig.\ \ref{fig12}b
($d\sigma / d\eta_{{\rm jet}2}$).  For $p_{T,{\rm jet}2} \ge 6$ GeV the
cross section is very much reduced as compared to the cross section in
Fig.\ \ref{fig11}a. In fact, the dominating event configuration is with
a photon and one jet balancing each other in $p_T$; a third jet with
comparable $p_T$ is found in only a small portion of the events.  The
rapidity distribution $d\sigma / d\eta_{{\rm jet}2}$ plotted in Fig.\ 
\ref{fig12}b has a larger tail extending to larger rapidities as
compared to the cross sections $d\sigma / d\eta_{\gamma}$ and $d\sigma /
d\eta_{{\rm jet}1}$.  The second energetic jet originates dominantly
from the incoming quark and therefore is in many cases closer to the
proton remnant, i.e.\ at positive rapidities, than the harder jet.

The cross sections $d\sigma / dp_{T,{\rm jet}2}$ and $d\sigma /
d\eta_{{\rm jet}2}$ increase with decreasing $R$. This is the behaviour
expected for cross sections which are of leading order in $\alpha_s$.

\begin{figure}[hb] 
\unitlength 1mm
\begin{picture}(160,78)
\put(40,-1){\epsfig{file=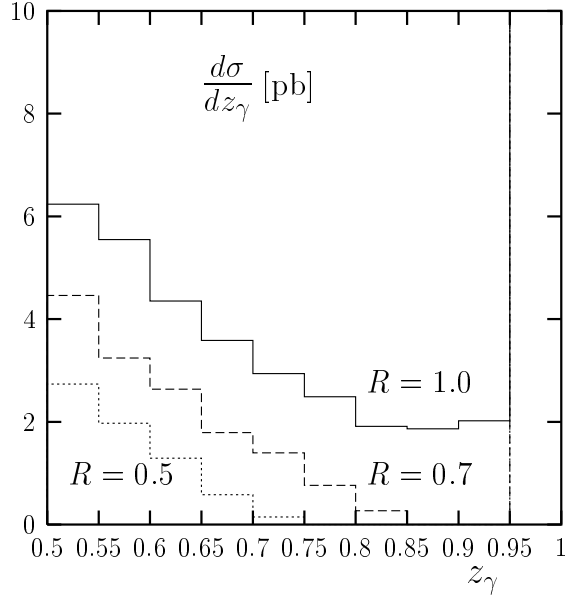,width=8cm}}
\end{picture}
\caption{$z_{\gamma}$ distributions for $R = 1.0$, 0.7 and 0.5 (full,
  dashed and dotted curves).}
\label{fig13}
\vspace*{-2.5cm}
\end{figure}

In our previous work \cite{GKS2} we studied $d\sigma / dz_{\gamma}$ as a
function of the fraction $z_{\gamma}$ of the momentum of the photon
inside the photon jet. This cross section is expected to contain
information on the photon fragmentation functions. The results presented
in \cite{GKS2} were obtained for the case $R = 1$ only. To see how
results change with $R$, we show in Fig.\ \ref{fig13} $d\sigma /
z_{\gamma}$ for the particular photon fragmentation function of Bourhis
et al.\ \cite{bfg} for $R = 0.5$, 0.7 and 1.0. As is seen, $d\sigma /
dz_{\gamma}$ decreases with decreasing $R$ since this cross section is a
superposition of leading and next-to-leading order contributions. 

\clearpage

The $R$ dependence of the other cross sections considered above follows
the same pattern. In cases where we have a superposition of LO and NLO
contributions we encounter a decreasing cross section with decreasing
$R$. In regions where the cross section receives contribution from
$O(\alpha_s)$ only with no additional NLO corrections included, the
cross section increases with decreasing $R$. Thus for example $d\sigma /
dz_{\gamma 1}$ (cf.\ Fig.\ \ref{fig7}a) is decreasing with $R$ in the
bin at $z_{\gamma 1} = 1$ and increasing with $R$ outside this bin,
i.e.\ for $z_{\gamma 1} < 0.9$ and $z_{\gamma 1} > 1.1$. The $R$
dependence of $d\sigma / dz_{12}$ (cf.\ Fig.\ \ref{fig7}b) is similar:
inside the bin around $z_{12} = 0$ the cross section decreases with
decreasing $R$ and outside $z_{12} = 0$ it increases with decreasing
$R$.

\subsection{Jet Algorithm Dependence}

\begin{figure}[t] 
\unitlength 1mm
\begin{picture}(160,80)
\put(0,-1){\epsfig{file=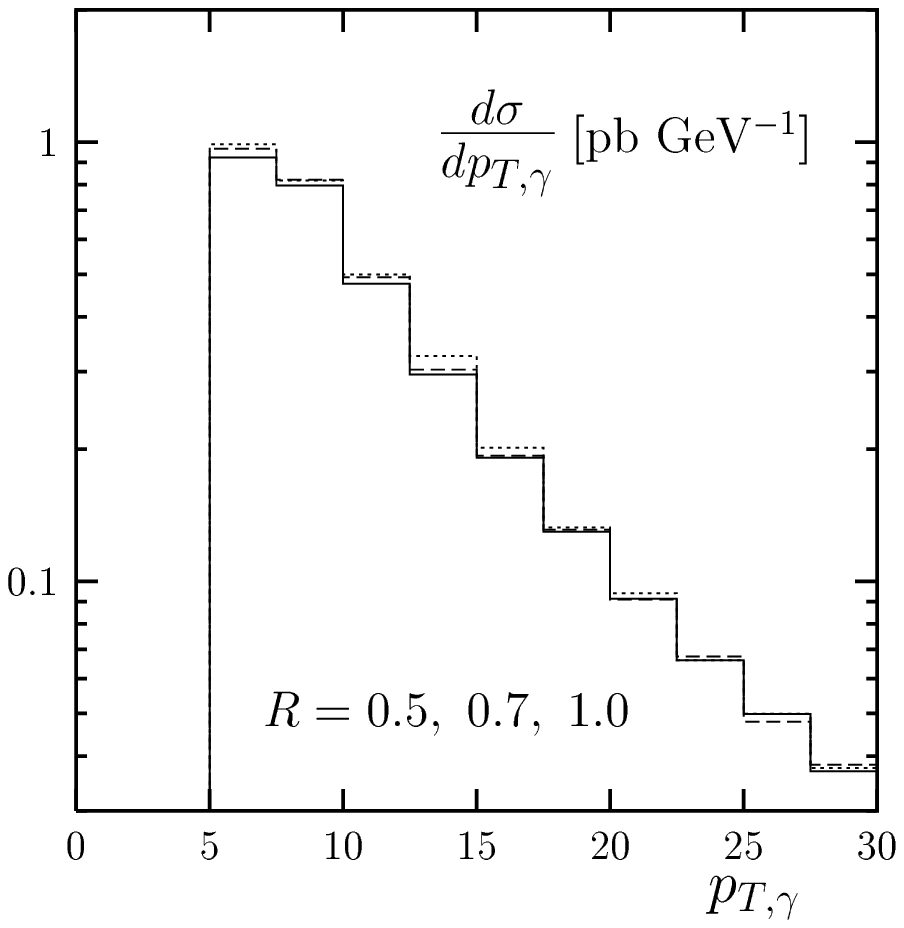,width=8cm}}
\put(40,-1){(a)}
\put(80,-1){\epsfig{file=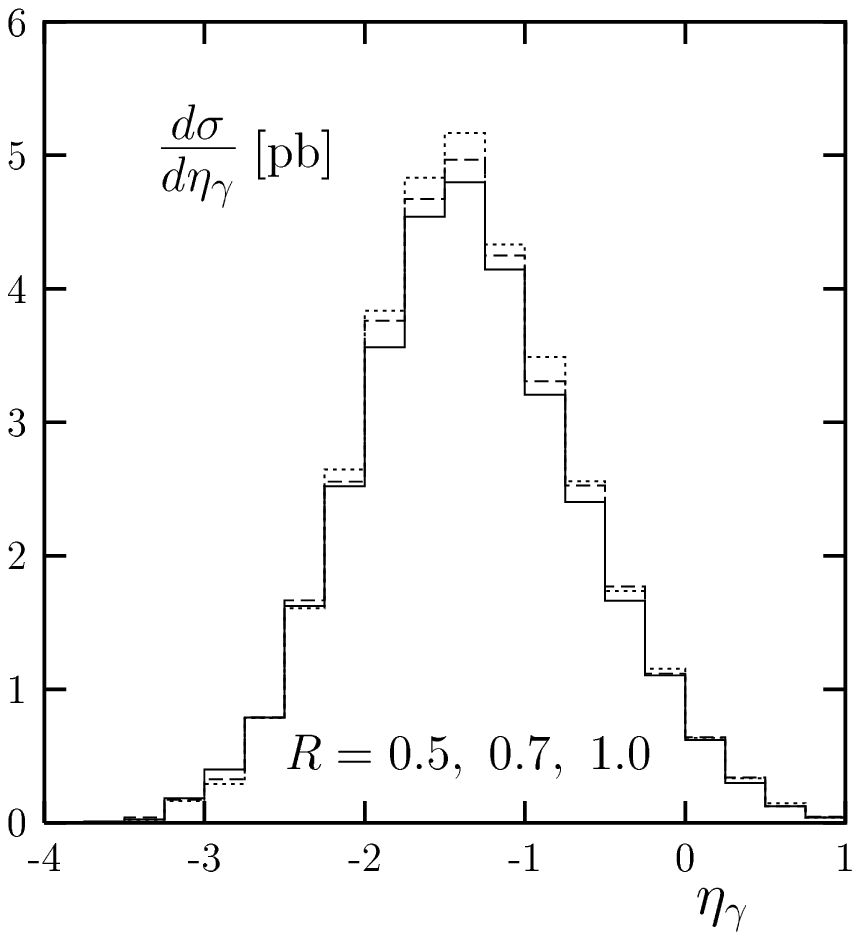,width=8cm}}
\put(120,-1){(b)}
\end{picture}
\caption{$p_T$- (a) and $\eta$- (b) distributions of the photon jet for
  $R = 1.0$, 0.7 and 0.5 (full, dashed and dotted curves) in the
  $k_T$ algorithm.} 
\label{fig15}
\end{figure}
\begin{figure}[t] 
\unitlength 1mm
\begin{picture}(160,80)
\put(0,-1){\epsfig{file=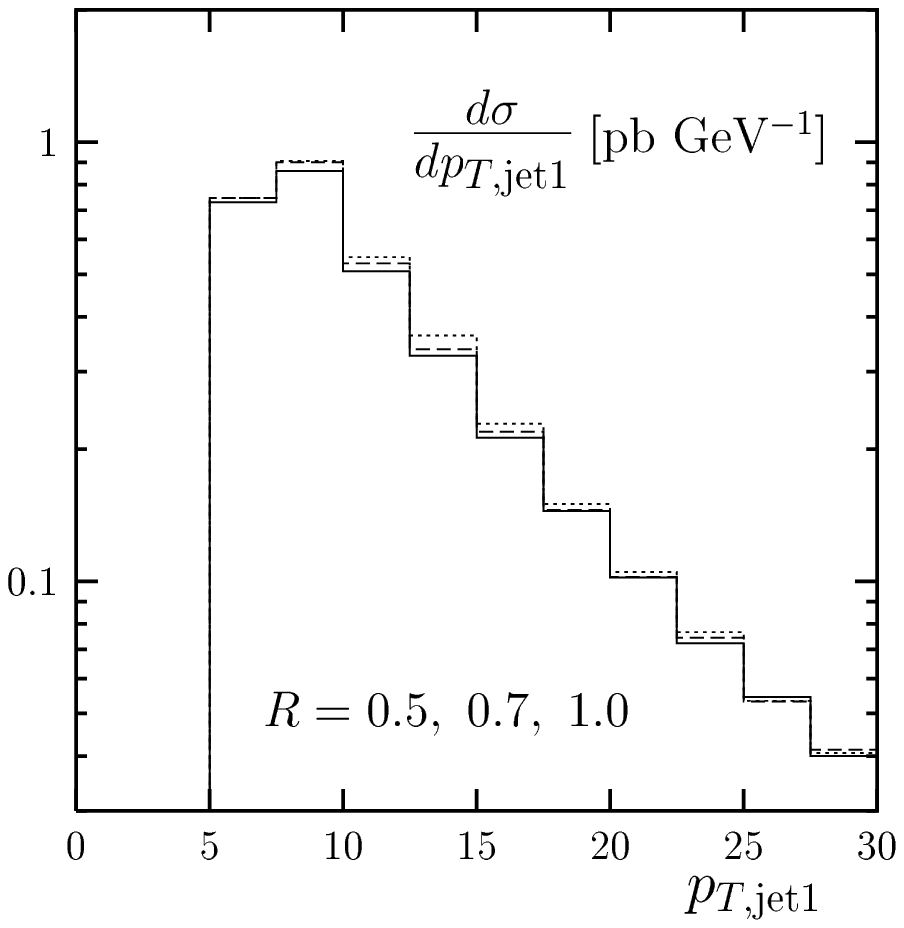,width=8cm}}
\put(40,-1){(a)}
\put(80,-1){\epsfig{file=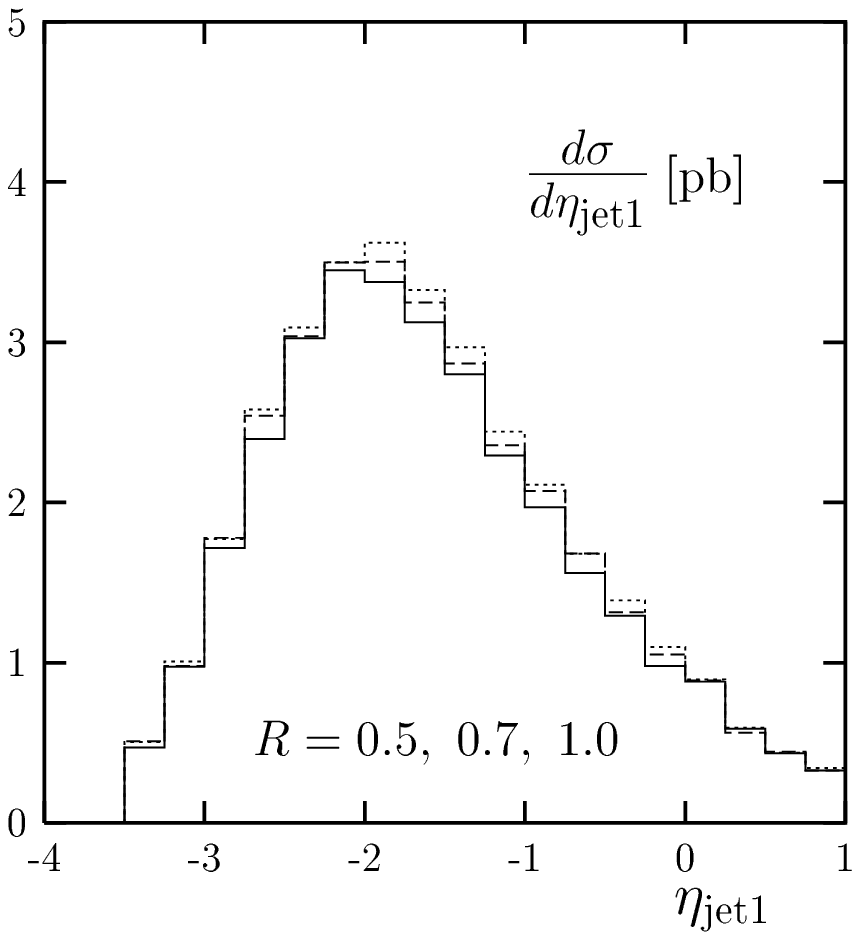,width=8cm}}
\put(120,-1){(b)}
\end{picture}
\caption{$p_T$- (a) and $\eta$- (b) distributions of the most energetic
  jet for $R = 1.0$, 0.7 and 0.5 (full, dashed and dotted curves) in the
  $k_T$ algorithm.} 
\label{fig16}
\end{figure}

\begin{figure}[t] 
\unitlength 1mm
\begin{picture}(160,80)
\put(0,-1){\epsfig{file=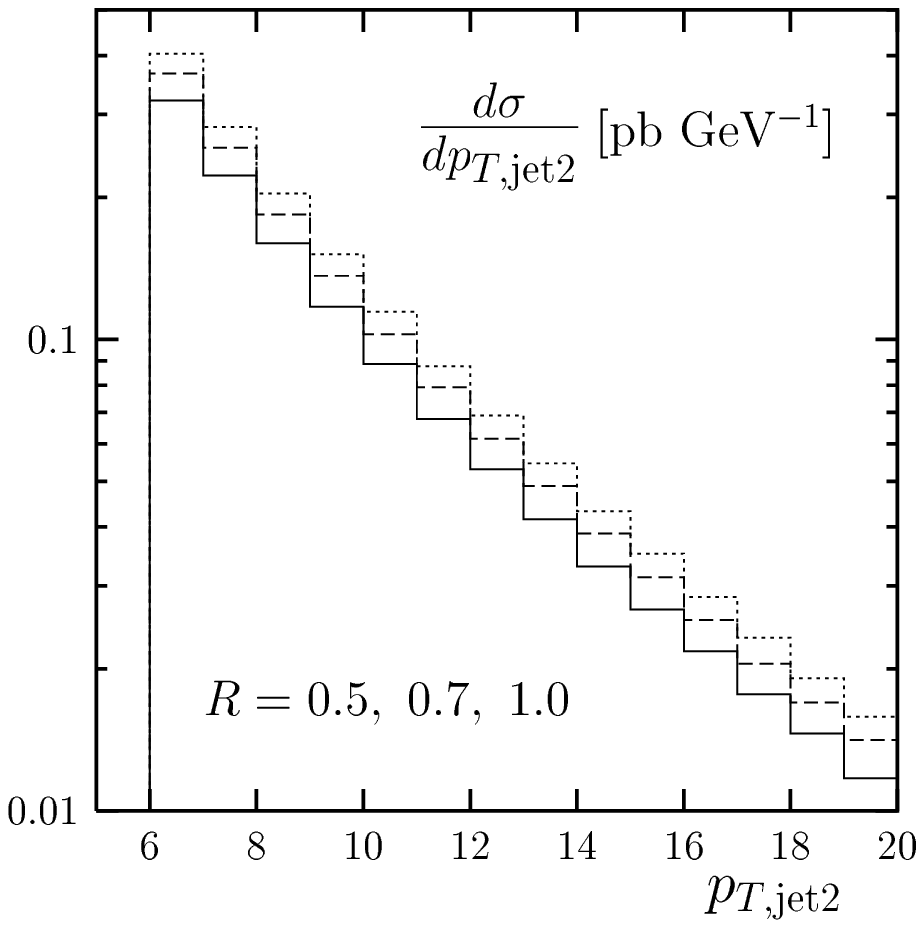,width=8cm}}
\put(40,-1){(a)}
\put(80,-1){\epsfig{file=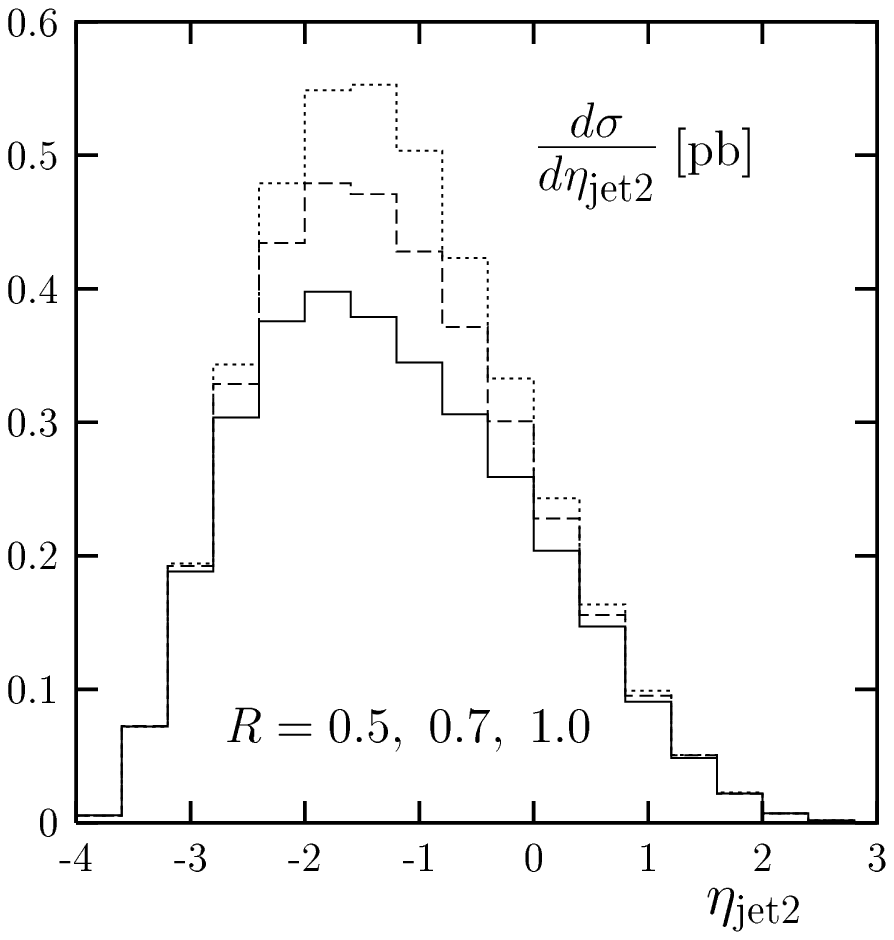,width=8cm}}
\put(120,-1){(b)}
\end{picture}
\caption{$p_T$- (a) and $\eta$-distributions (b) of the second jet for
  $R = 1.0$, 0.7 and 0.5 (full, dashed and dotted curves) in the
  $k_T$ algorithm.} 
\label{fig17}
\end{figure}

Measurements of cross sections for the production of a photon plus jets,
as for example in $\gamma p$ collisions \cite{HERA-data}, have been
performed with the help of the cone algorithm used to define jets and
the photon isolation. However, the $k_T$ algorithm has definite
advantages, in particular in the experimental analysis \cite{seymour}.
Therefore we present a few distributions based on the $k_T$ algorithm as
well. This algorithm is used for the recombination of two partons into a
jet as explained in sect.\ 3.1 as well as for the definition of the
photon jet. In Figs.\ \ref{fig15}a, \ref{fig16}a we show $d\sigma /
dp_{T,\gamma}$ and $d\sigma / dp_{T,{\rm jet}1}$ calculated with the
$k_T$ algorithm for $R = 0.5$, 0.7 and 1. These two groups of curves
have to be compared to the results with the cone algorithm in Figs.\ 
\ref{fig10}a, \ref{fig11}a. The qualitative behaviour of the two cross
sections is similar; but there are quantitative differences. We note
that the dependence on the parameter $R$, which controls the size of the
jets, is now even more reduced as compared to the corresponding cross
sections with the cone algorithm. The cross sections $d\sigma /
d\eta_{\gamma}$ and $d\sigma / d\eta_{{\rm jet}1}$ for the $k_T$
algorithm are displayed in Figs.\ \ref{fig15}b, \ref{fig16}b. They can
be compared with the corresponding cross section for the cone algorithm
in Figs.\ \ref{fig10}b, \ref{fig11}b. Similar to the $p_T$ distributions
the qualitative behaviour did not change.  Again the $R$ dependence
seems to be reduced for the $k_T$ algorithm.  This is even more the case
for the cross sections $d\sigma / dp_{T,{\rm jet}2}$ and $d\sigma /
d\eta_{{\rm jet}2}$, shown in Figs.\ \ref{fig17}a, b which should be
compared to the corresponding results in Figs.\ \ref{fig12}a, b. It is
clear that the cross sections in Figs.\ \ref{fig17}a, b increase with
decreasing $R$ in the same way as the cross sections for the cone
algorithm in Figs.\ \ref{fig12}a, b.

\begin{figure}[b] 
\unitlength 1mm
\begin{picture}(160,70)
\put(0,-1){\epsfig{file=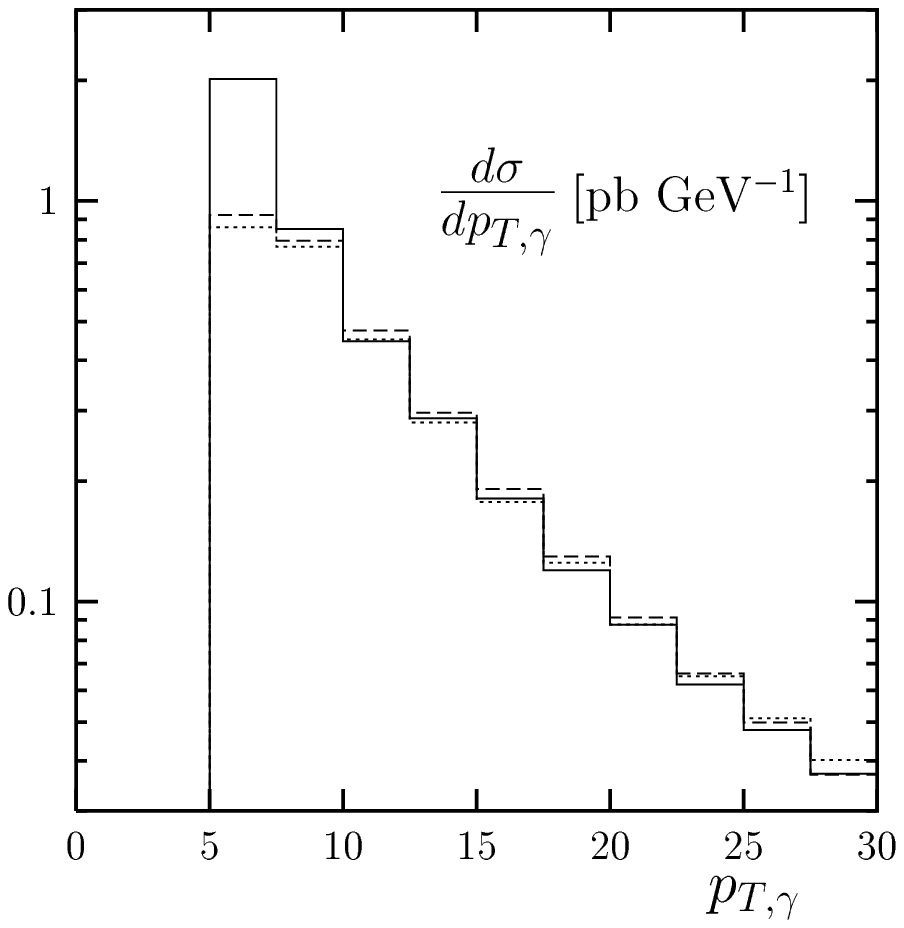,width=8cm}}
\put(40,-1){(a)}
\put(80,-1){\epsfig{file=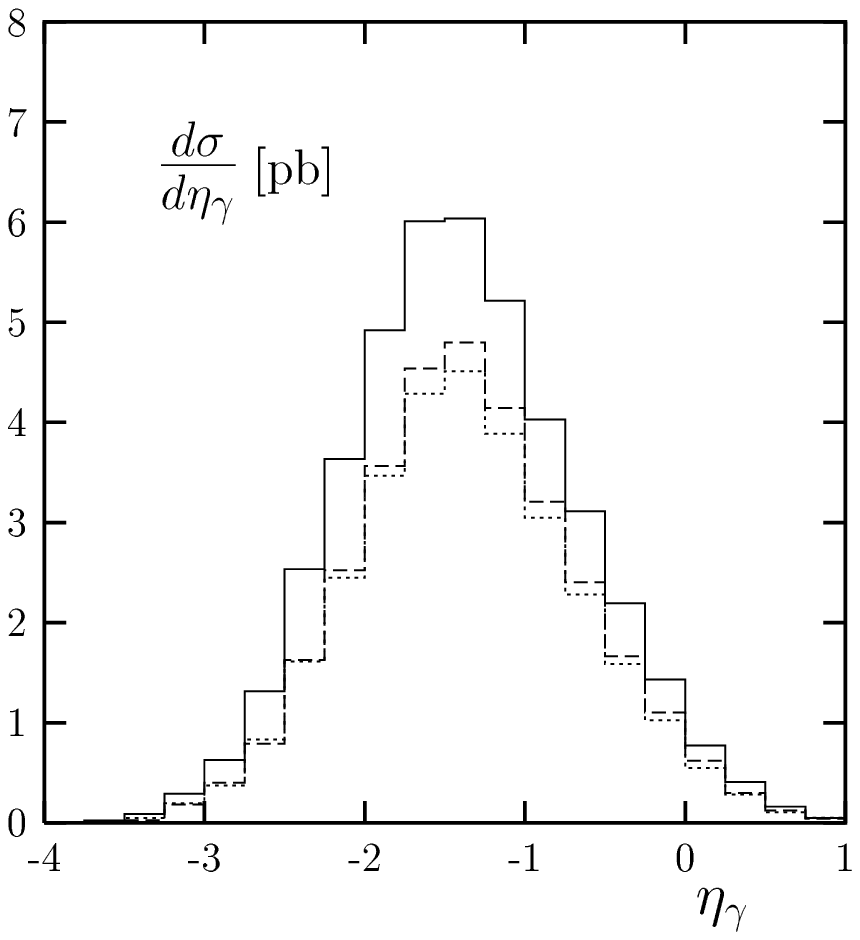,width=8cm}}
\put(120,-1){(b)}
\end{picture}
\caption{Comparison of the $p_T$- (a) and $\eta$-distributions (b) of
  the photon jet in the cone algorithm with $z_{\gamma} \geq 0.9$ and
  $R=1.0$ (dotted curves), in the $k_T$ algorithm with $R = 1.0$ (dashed
  curves) and without jet analysis (inclusive cross section, full
  curves).} 
\label{fig18}
\end{figure}

A direct comparison of cross sections calculated with either the cone
algorithm or the $k_T$ algorithm is shown in Fig.\ \ref{fig18} (dashed
and dotted curves). Here we have chosen $R=1$ and $z_{\rm cut} = 0.9$.
In Figs.\ \ref{fig18}a, b the cross sections $d\sigma / dp_{T,\gamma}$
and $d\sigma / d\eta_{\gamma}$ are plotted, respectively. As we can see,
these cross sections hardly change when the cone algorithm is replaced
by the $k_T$ algorithm. Only where $d\sigma / d\eta_{\gamma}$ is
maximal, i.e.\ near $\eta_{\gamma} = -1.5$, the cross section with the
$k_T$ algorithm is approximately 5\,\% larger than with the cone
algorithm.

\clearpage

It is expected that the requirement to observe additional jets reduces
the cross section for the production of a high-$p_T$ photon. To study
this reduction we have calculated also $d\sigma / dp_{T,\gamma}$ and
$d\sigma / d\eta_{\gamma}$ for the inclusive case, i.e.\ without
additional jets required in the final state and the same photon
isolation constraint in terms of an isolation cone around the photon and
the cut $z_{\rm cut} = 0.9$. The results for these inclusive cross
sections are shown in Figs.\ \ref{fig18}a, b as full curves. The total
cross section without jet algorithm applied is increased by about
35\,\%. This is reflected in the inclusive cross section $d\sigma /
d\eta_{\gamma}$ which is always larger than the corresponding
distribution for final states which are required to contain at least one
jet, over the full range of $\eta_{\gamma}$. In $d\sigma /
dp_{T,\gamma}$ only the first two $p_T$-bins contain an appreciably
larger cross section due to the removal of the jet requirement. This can
be traced back to the different prescriptions used to remove events with
small transverse momentum: for the analyses based on the cone or the
$k_T$ jet algorithms, we used the $p_T$ cuts (\ref{ptcuts}) which are
applied to individual jets, whereas in the case without jet algorithm we
applied the cut (\ref{ETcuts}) to the sum of transverse momenta of all
hadronic particles in the final state. As a consequence, an event with
two low-$p_T$ partons, each with $p_T < 6$ GeV will be rejected in the
first case if these two partons are not recombined into one jet,~ while
the event will be accepted in the second case if the sum of the
transverse momenta of the two partons is larger than the cut of 6 GeV.
This clearly affects only the bins with lowest $p_T$.

\subsection{Scale Dependence of Jet Cross Sections}

All results presented so far have been obtained for a renormalization
and factorization scale $\mu$ fixed at $\mu = p_{T,{\rm jet}1}$ ($\mu =
E_T$ of (\ref{ETcuts}) for inclusive cross sections). In LO cross
sections, the scale dependence is exclusively due to variations of $\mu$
in the parton distribution functions. At NLO we expect that additional
terms containing an explicit $\mu$-dependence will reduce the scale
dependence.  Instead of studying the scale dependence of all the
differential cross sections discussed so far separately, we have
investigated the scale dependence of some components of the total cross
section, i.e.\ integrated over the phase space allowed by the transverse
momentum cuts (\ref{ptcuts}). We define the scale in the form $\mu^2 =
f^2 p^2_{T,{\rm jet}1}$ and vary $f$ between $f = 1/4$ and 4.

\begin{figure}[tb] 
\unitlength 1mm
\begin{picture}(160,80)
\put(0,-1){\epsfig{file=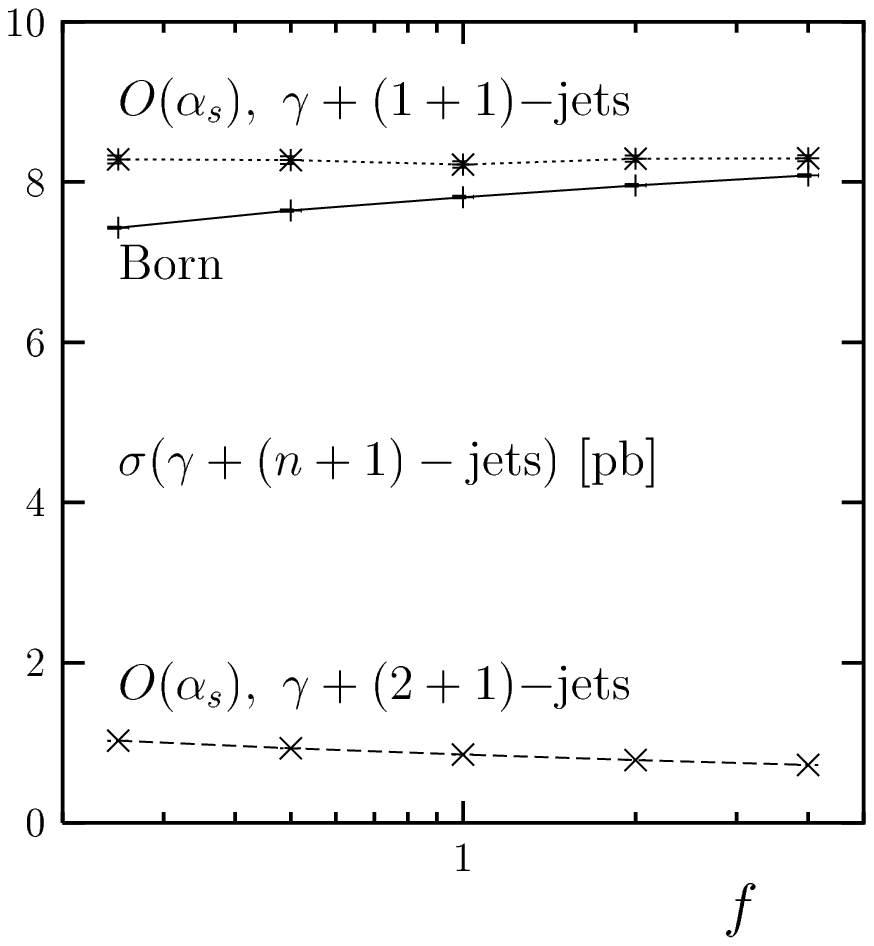,width=8cm}}
\put(38,-1){(a)}
\put(80,-1){\epsfig{file=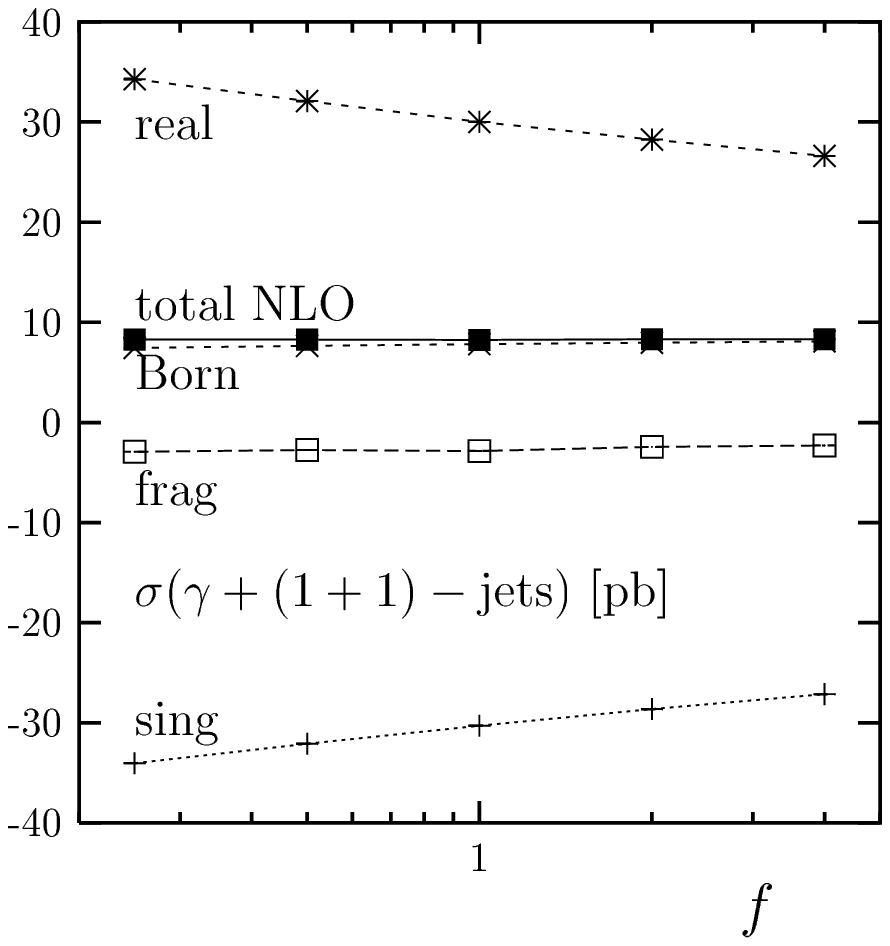,width=8cm}}
\put(118,-1){(b)}
\end{picture}
\caption{Scale dependence of the total cross section (a) and of separate 
  contributions to the $\gamma + (1+1)$-jet cross section (b).}
\label{fig14}
\end{figure}

In Fig.\ \ref{fig14}a we have plotted the $f$ dependence of the $\gamma
+ (1+1)$-jet cross section in LO and NLO (denoted $O(\alpha_s)$) and of
the $\gamma + (2+1)$-jet cross section. The LO cross section (denoted
``Born'' in Fig.\ \ref{fig14}) increases with $f$ by approximately
10\,\% in the range $0.25 < f < 4$. The cross section including
corrections of $O(\alpha_s)$ is almost independent of $f$, i.e.\ no
scale dependence inside the considered range of scales is visible. Here
the decrease of the cross section due to the decrease of $\alpha_s$ with
increasing $f$ is compensated by the increase originating from the scale
dependence of the proton PDF's. The cross section for the $\gamma +
(2+1)$-jet final state, being of order $O(\alpha_s)$, decreases with
increasing $f$ by approximately 25\,\%. This is a combined effect of the
dependence of $\alpha_s$ and of the $f$ dependence of the parton
distribution functions. The $f$ dependence of separate components
(``real'', ``sing'' and ``frag'') of the $\gamma + (1+1)$-jet cross
section including $O(\alpha_s)$ terms is plotted in Fig.\ \ref{fig14}b
and again compared with the LO cross section. ``real'' stands for the
tree graph contributions in $O(\alpha_s)$, calculated with the slicing
parameter as described in section 2.4.  Since it is a tree-graph term it
decreases with increasing $f$ due to the decrease of $\alpha_s$.  The
singular term ``sing'', which includes virtual corrections and singular
contributions below the slicing cut, is negative and decreases in
absolute value by the same amount as ``real''.  The fragmentation
contribution ``frag'' which includes both terms of the right-hand side
of (\ref{dqg}) is also negative and almost independent of $f$ as
expected, since the factorization scale dependence ($\mu_F$ in
(\ref{dqg})) cancels in first approximation.

\subsection{Dependence on Low-$p_T$ Cuts}

The choice of two different cuts for the transverse momentum of the
photon and the jet is needed to avoid the otherwise present infrared
sensitivity of the NLO predictions. This sensitivity is known from
similar calculations of dijet cross sections in $ep$ collisions
\cite{kpoet} and must be avoided. The same problem was encountered in
the calculation of inclusive two-jet cross sections in $\gamma p$
collisions \cite{kkfr}, for the production of a prompt photon plus a
charm quark in $p\bar{p}$ collisions \cite{bailey} and much earlier in
NLO calculations of the inclusive cross section for photon-hadron
\cite{aurenche} and for two-photon production \cite{adbfs2}.

\begin{figure}[tb] 
\unitlength 1mm
\begin{picture}(160,80)
\put(0,-1){\epsfig{file=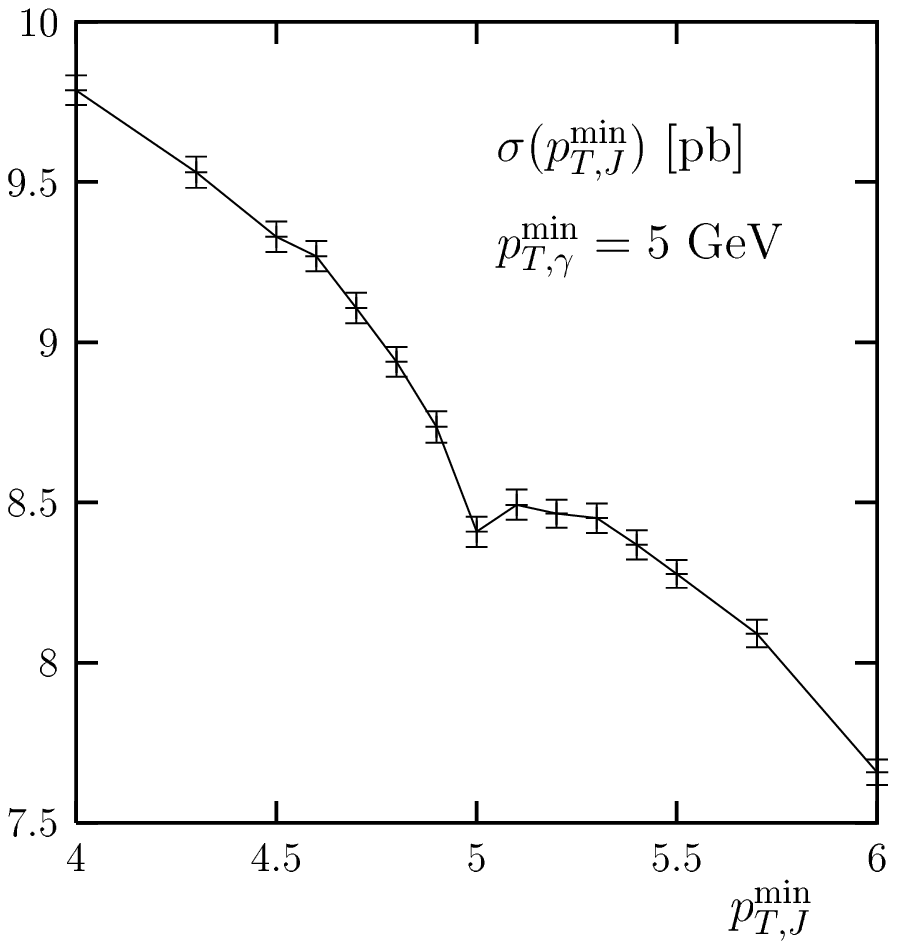,width=8cm}}
\put(38,-1){(a)}
\put(80,-1){\epsfig{file=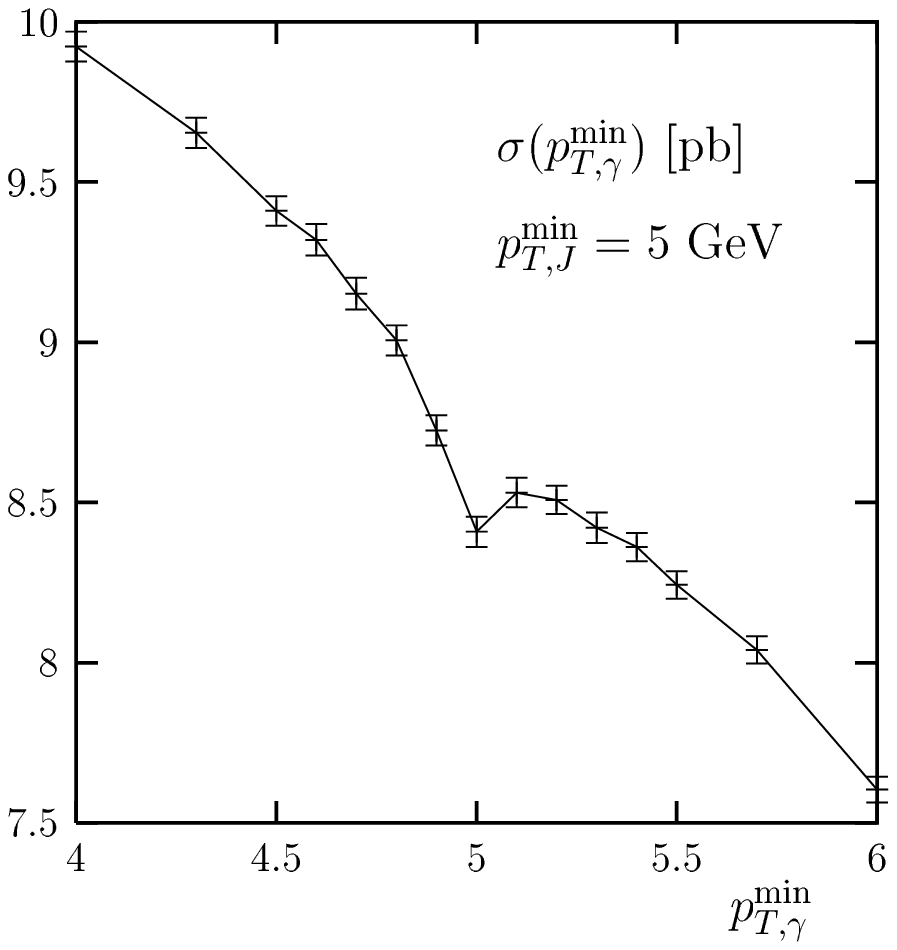,width=8cm}}
\put(118,-1){(b)}
\end{picture}
\caption{Dependence of the total cross section on lower transverse
  momentum cutoffs: as a function of $p_{T,J}^{\rm min}$ with fixed
  $p_{T,\gamma}^{\rm min} = 5$ GeV (a) and as a function of
  $p_{T,\gamma}^{\rm min}$ with fixed $p_{T,J}^{\rm min} = 5$ GeV (b).} 
\label{fig20}
\end{figure}

Above we have chosen the difference between the two $p_{T,{\rm min}}$
cuts for the photon and the jet, $\Delta = p_{T,J}^{\rm min} -
p_{T,\gamma}^{\rm min}$, equal to 1 GeV (see (\ref{ptcuts})). $\Delta$
should not be too small since then we would encounter the infrared
sensitive region where the prediction of the cross section becomes
unreliable. In order to obtain some information about possible choices
for $\Delta$ we have studied the $\gamma + (n+1)$-jet cross section
$d\sigma / dp_{T,J}$ integrated over $p_{T,J} \geq p_{T,J}^{\rm min}$ as
a function of $p_{T,J}^{\rm min}$. The transverse momentum of the photon
was always integrated over the range $p_{T,\gamma} \geq
p_{T,\gamma}^{\rm min} = 5$ GeV. The results for $\sigma(p_{T,J}^{\rm
  min})$ are plotted in Fig.\ \ref{fig20}a. Starting at $p_{T,J}^{\rm
  min} = 6$ GeV this cross section increases with decreasing
$p_{T,J}^{\rm min}$ with almost constant slope. At about $p_{T,J}^{\rm
  min} = 5.5$ GeV the slope decreases and approaches zero and even
changes sign so that $\sigma(p_{T,J}^{\rm min})$ develops a maximum
below $p_{T,J}^{\rm min} = 5.5$ GeV. This change of slope is due to the
infrared sensitivity in the point $p_{T,J}^{\rm min} = p_{T,\gamma}^{\rm
  min}$.  To avoid this region one must choose $\Delta \ne 0$. From the
plot we observe that $\Delta \geq 0.5$ GeV would be sufficient. Thus, in
principle, we could have used a smaller value for this difference than
was chosen in (\ref{ptcuts}).

The cross section in the vicinity of $p_{T,J}^{\rm min} =
p_{T,\gamma}^{\rm min}$ cannot be predicted reliably. It depends on the
technical cut $y_0^J$ as soon as one approaches the limit $\Delta =
0$. In fact, the infrared sensitive region is very much influenced by
non-perturbative effects which are not included in our calculation. In
any case it would be interesting to measure $\sigma(p_{T}^{\rm min})$ in 
order to investigate this non-perturbative region. 

In Fig.\ \ref{fig20}a we present $\sigma(p_{T,J}^{\rm min})$ also for
$p_{T,J}^{\rm min}$ below 5 GeV, i.e.\ for $\Delta < 0$. If
$p_{T,J}^{\rm min}$ is increased starting from $p_{T,J}^{\rm min} = 4$
GeV, the slope of $\sigma(p_{T,J}^{\rm min})$ changes at about
$p_{T,J}^{\rm min} = 4.5$ GeV and $\sigma(p_{T,J}^{\rm min})$ becomes
smaller towards $p_{T,J}^{\rm min} = 5$ GeV compared to the behaviour
with constant slope. This stronger decrease of $\sigma(p_{T,J}^{\rm
  min})$ above $p_{T,J}^{\rm min} = 4.5$ has its origin again in the
infrared sensitivity of this region. For $p_{T,J}^{\rm min} \rightarrow
5$ GeV, and $\Delta < 0$ the cross section approaches the same value as
we have obtained for $\Delta \rightarrow 0$ in the region $\Delta > 0$.
This means, $\sigma(p_{T,J}^{\rm min})$ is not singular at $p_{T,J}^{\rm
  min} = 5$ GeV, nor has it a discontinuity.

The observed behaviour is of course not only visible at the specific
value of $p_T^{\rm min} = 5$ GeV which we have chosen. A similar dip is
seen for larger and smaller values as well. It becomes more pronounced
for smaller values and gets washed out for larger $p_T^{\rm min}$. 

For completeness we show in Fig.\ \ref{fig20}b also the cross section
with the roles of $p_{T,J}^{\rm min}$ and $p_{T,\gamma}^{\rm min}$
interchanged, i.e.\ we fix $p_{T,J}^{\rm min} = 5$ GeV and study the
dependence on the cut for the photon transverse momentum
$p_{T,\gamma}^{\rm min}$. The behaviour of the cross section in Fig.\ 
\ref{fig20}b is very similar to the first case and shows the same
infrared sensitive region. Inside the region $-0.5$ GeV $< \Delta <
0.5$ GeV the cross sections agree inside numerical errors. Only for
larger $|\Delta|$ the dependences on the minimal transverse momenta of
the jet and the photon become different. The cross sections shown in
Fig.\ \ref{fig20} have been calculated with the cone algorithm using $R
= 1$ and $z_{\rm cut} = 0.9$.

\section{Summary and Concluding Remarks}

We have presented results of a next-to-leading order calculation of
isolated photon production in large-$Q^2$ $ep$ scattering. Contributions 
from quark-to-photon fragmentation are explicitly taken into account. We 
have discussed numerical results for $\gamma + (1+1)$-jet and $\gamma +
(2+1)$-jet cross sections as functions of transverse momenta, rapidity
and other observables derived from photon and/or jet kinematic
variables. Infrared sensitive regions, as for example the region of
equal photon and jet $p_T$, are studied in detail. 

We investigated several of these cross sections with respect to their
photon isolation, jet cone size, scale and jet algorithm dependences. It
was found that these dependences are rather weak. In particular, the
scale dependence of the integrated cross section is very small, giving
quite some confidence in the reliability of our predictions. Also the
results depended very little on the choice of modern parton distribution
functions for the proton.

We expect that the measurement of photon plus jet cross sections at HERA
will contribute to testing perturbative QCD in the process
$\gamma^{\ast} p \rightarrow \gamma X$, in an area which has not been
studied yet. The calculation covers the range of large $Q^2$ where the
results do not depend on parton distribution functions of the virtual
photon. At even larger $Q^2$ additional contributions from $Z$ boson
exchange become important, which have been neglected in the present
calculation. It is no major problem to incorporate these missing parts.
They are non-negligible and must be considered when experimental data
become available at very large $Q^2$.



\begin{thebibliography}{99}
  
\bibitem{apana} For a recent discussion see: L.\ Apanasevich et al.,
  Phys.\ Rev.\ {\bf D59} (1999) 074007, and the earlier references given
  there.

\bibitem{HERA-data} J.\ Breitweg et al., ZEUS Collaboration, Phys.\
  Lett.\ {\bf B413} (1997) 201; 
  K.\ M\"uller, Proceedings of the Eur.\ Phys.\ Soc.\ {\it High Energy
    Physics Conference}, Jerusalem 1997, p.\ 464; 
  J.\ Breitweg et al., ZEUS Collaboration, DESY-99-161, October 1999,
  (to appear in Phys.\ Lett.\ B).

\bibitem{aurenche} P.\ Aurenche, R.\ Baier, A.\ Douiri, M.\ Fontannaz
  and D.\ Schiff, Z.\ Phys.\ {\bf C24} (1984) 309.
 
\bibitem{koller}  K.\ Koller, T.\ F.\ Walsh and P.\ M.\ Zerwas, Z.\
  Phys.\ {\bf C2} (1979) 197. 

\bibitem{ppbar} H.\ Baer, J.\ Ohnemus and F.\ F.\ Owens, Phys.\ Rev.\
  {\bf D41} (1990) 61;
  P.\ Aurenche et al., Nucl.\ Phys.\ {\bf B399} (1993) 34;
  L.\ E.\ Gordon and W.\ Vogelsang, Phys.\ Rev.\ {\bf D50} (1994) 1901;
  L.\ E.\ Gordon, E.\ Reya and W.\ Vogelsang, Phys.\ Rev.\ Lett.\ {\bf
    73} (1994) 388.

\bibitem{e+e-} G.\ Kramer and B.\ Lampe, Phys.\ Lett.\ {\bf B269} (1991) 
  401;
  G.\ Kramer and H.\ Spiesberger, Proceedings {\it Workshop on Photon
    Radiation from Quarks}, Annecy 1991, CERN Yellow Report 92-04, p.\
  26; 
  E.\ W.\ N.\ Glover and W.\ J.\ Stirling, Phys.\ Lett.\ {\bf B295}
  (1992) 128; 
  Z.\ Kunszt and Z.\ Tr\'ocs\'anyi, Nucl.\ Phys.\ {\bf B394} (1993)
  139; 
  E.\ L.\ Berger, X.\ Guo and J.\ Qiu, Phys.\ Rev.\ {\bf D54} (1996)
  5470. 

\bibitem{GGdR1} A.\ Gehrmann-De Ridder, T.\ Gehrmann and E.\ W.\ N.\
  Glover, Phys.\ Lett.\ {\bf B414} (1997) 354. 

\bibitem{GGdR2} A.\ Gehrmann-De Ridder and E.\ W.\ N.\ Glover, Nucl.\
  Phys.\ {\bf B517} (1998) 269; Eur.\ Phys.\ J.\ {\bf C7} (1999) 29.

\bibitem{phopro} P.\ Aurenche et al., Z.\ Phys.\ {\bf C56} (1993) 589; 
  L.\ E.\ Gordon and J.\ K.\ Storrow, Z.\ Phys.\ {\bf C63} (1994) 581; 
  L.\ E.\ Gordon and W.\ Vogelsang, Phys.\ Rev.\ {\bf D52} (1995) 58; 
  L.\ E.\ Gordon, Phys.\ Rev.\ {\bf D57} (1998) 235; 
  M.\ Krawczyk and A.\ Zembrzuski, Proceedings of the {\em 29th
    International Conference on High-Energy Physics (ICHEP 98)},
  Vancouver, Canada, 23-29 July 1998, Vol.\ 1, p.\ 895,
  (hep-ph/9810253).   

\bibitem{KMS} G.\ Kramer, D.\ Michelsen and H.\ Spiesberger, Eur.\
  Phys.\ J.\ {\bf C5} (1998) 293.

\bibitem{GKS} A.\ Gehrmann-De Ridder, G.\ Kramer and H.\ Spiesberger,
  Phys.\ Lett.\ {\bf B459} (1999) 271.

\bibitem{GKS2} A.\ Gehrmann-De Ridder, G.\ Kramer and H.\ Spiesberger,
  Eur.\ Phys.\ J.\ {\bf C11} (1999) 137.

\bibitem{dmich} D.\ Michelsen, doctoral thesis, University of Hamburg,
  1995, DESY-95-146.

\bibitem{slicing} K.\ Fabricius, G.\ Kramer, G.\ Schierholz and I.\
  Schmidt, Z.\ Phys.\ {\bf C11} (1982) 315; 
  F.\ Gutbrod, G.\ Kramer and G.\ Schierholz, Z.\ Phys.\ {\bf C21}
  (1984) 235; 
  W.\ T.\ Giele and E.\ W.\ N.\ Glover, Phys.\ Rev.\ {\bf D46} (1992)
  1980. 

\bibitem{glover1} E.\ W.\ N.\ Glover and A.\ G.\ Morgan, Z.\ Phys.\ {\bf 
    C62} (1994) 311. 

\bibitem{mrs} A.\ D.\ Martin, R.\ G.\ Roberts, W.\ J.\ Stirling and R.\
  S.\ Thorne, Eur.\ Phys.\ J.\ {\bf C4} (1998) 463.

\bibitem{mrs99} A.\ D.\ Martin, R.\ G.\ Roberts, W.\ J.\ Stirling and
  R.\ S.\ Thorne, Univ.\ Durham preprint DTP/99/64 (1999),
  (hep-ph/9907231). 

\bibitem{cteq4+5} H.\ L.\ Lai et al., Phys.\ Rev.\ {\bf D55} (1997); 
  H.\ L.\ Lai and W.\ K.\ Tung, Z.\ Phys.\ {\bf C74} (1997) 463; 
  H.\ L.\ Lai et al., Eur.\ Phys.\ J.\ {\bf C12} (2000) 375. 

\bibitem{bfg} L.\ Bourhis, M.\ Fontannaz and J.\ Ph.\ Guillet, Eur.\
  Phys.\ J.\ {\bf C2} (1998) 529.

\bibitem{aleph} D.\ Buskulic et al., ALEPH Collaboration, Z.\ Phys.\
  {\bf C69} (1996) 365.

\bibitem{opal} K.\ Ackerstaff et al., OPAL Collaboration, Eur.\ Phys.\
  J.\ {\bf C2} (1998) 39.

\bibitem{seymour} M.\ H.\ Seymour, Nucl.\ Phys.\ {\bf B513} (1998) 269. 

\bibitem{kt-algorithm} S.\ Catani, Y.\ L.\ Dokshitzer, M.\ H.\ Seymour
  and B.\ R.\ Webber, Nucl.\ Phys.\ {\bf B406} (1993) 187; 
  S.\ D.\ Ellis and D.\ E.\ Soper, Phys.\ Rev.\ {\bf D48} (1993) 3160. 

\bibitem{bailey} B.\ Bailey, E.\ L.\ Berger and L.\ E.\ Gordon, Phys.\
  Rev.\ {\bf D54} (1996) 1896; 
  E.\ L.\ Berger and L.\ E.\ Gordon, Phys.\ Rev.\ {\bf D54} (1996)
  2279. 

\bibitem{kpoet} G.\ Kramer and B.\ P\"otter, Eur.\ Phys.\ J.\ {\bf C5}
  (1998) 665. 

\bibitem{kkfr} M.\ Klasen and G.\ Kramer, Phys.\ Lett.\ {\bf B366}
  (1996) 385; 
  S.\ Frixione and G.\ Ridolfi, Nucl.\ Phys.\ {\bf B507} (1997) 315. 

\bibitem{adbfs2} P.\ Aurenche, A.\ Douiri, R.\ Baier, M.\ Fontannaz and 
  D.\ Schiff, Z.\ Phys.\ {\bf C29} (1985) 459. 

\end{thebibliography}
\end{document}